\documentclass[12pt]{article}
\usepackage{epsfig,amsfonts,amssymb}
\usepackage{hyperref}
\pdfoutput=1

\usepackage{cite}
\usepackage{cite}

\usepackage{comment}

\def\ZZZ{{\hbox{ Z\kern-1.6mm Z}}}
\def\RRR{{\hbox{ R\kern-2.4mm R}}}
\def\CCC{{\hbox{ C\kern-2.0mm C}}}
\def\zzz{{\hbox{z\kern-1mm z}}}

\def\ZZZ{\mathbb{Z}}
\def\RRR{\mathbb{R}}

\newcommand{\qeq}{{\hbox{=\kern-2.3mm ? \kern.5mm }}}
\renewcommand{\qeq}{=}

\newcommand{\KK}{{\cal K}}

\newcommand{\CC}{{\cal C}}

\newcommand{\OO}{{\cal O}}

\newcommand{\wt}{\widetilde}

\newcommand{\RR}{{\cal R}}
\newcommand{\NN}{{\cal N}}
\newcommand{\TT}{{\cal T}}

\newcommand{\be}{\begin{equation}}
\newcommand{\ee}{\end{equation}}
\newcommand{\ben}{\begin{eqnarray}\displaystyle}
\newcommand{\een}{\end{eqnarray}}

\newcommand{\refb}[1]{(\ref{#1})}
\newcommand{\p}{\partial}
\newcommand{\sectiono}[1]{\section{#1}\setcounter{equation}{0}}

\def\one{{\hbox{ 1\kern-.8mm l}}}
\def\zero{{\hbox{ 0\kern-1.5mm 0}}}

\newcommand{\bea}[1]{\begin{eqnarray}\label{#1} }
\newcommand{\eea}{\end{eqnarray}}

\newcommand{\eqref}{\refb}

\usepackage{bm}
\usepackage[table]{xcolor}

\begin{document}

\baselineskip 24pt

\begin{center}

{\Large \bf Exotic Branes and Symmetries of String Theory}

\end{center}

\vskip .6cm
\medskip

\vspace*{4.0ex}

\baselineskip=18pt

\centerline{\large \rm Ashoke Sen}

\vspace*{4.0ex}

\centerline{\large \it International Centre for Theoretical Sciences - TIFR 
}
\centerline{\large \it  Bengaluru - 560089, India}


\vspace*{1.0ex}
\centerline{\small E-mail:  ashoke.sen@icts.res.in}

\vspace*{5.0ex}

\centerline{\bf Abstract} \bigskip

Are duality transformations symmetries of string theory? For AdS space-time the answer is 
no for generic asymptotic values of the moduli, since the duality symmetry is broken explicitly in the dual conformal field theory. In contrast, in string theory in flat space-time, monodromies around codimension two exotic branes  show that 
duality transformations are spontaneously broken discrete gauge symmetries with observable
consequences,  provided macroscopic loops of these branes  are not hidden 
behind an event horizon. 
We discuss how this can be achieved and how the situation in flat space-time
differs from that in AdS space-time. We also discuss observability of codimension two
non-BPS branes, codimension one BPS and non-BPS branes and higher codimension branes
of infinite tension.

\vfill \eject

\tableofcontents

\sectiono{Introduction} \label{s1}

String theory contains branes of different dimensions.
Since these are extended objects, they have infinite energy and are not regular states
of the theory. However one can construct closed loops of these branes by taking the spatial
part of the world-volume of a
$p$-brane to lie along a sphere $S^p$ or any other compact $p$-dimensional subspace and 
get a finite energy state of the theory.
Furthermore, as long as the size of $S^p$ is large, this
configuration will locally look like a flat $p$-brane and an asymptotic observer may be able
to study its properties. This however requires that the configuration is not hidden from the
asymptotic observer by the event horizon of a black hole.

For branes of codimension larger than two,  there is no problem in producing a
macroscopic loop that is not shielded by an event horizon. 
If we use Planck units to express all quantities, then 
for a codimension $k$ brane of
tension $\TT$ in $D$ space-time dimensions, a loop of size $L$ will have mass $M\sim
\TT\, L^{D-k-1}$
and hence its Schwarzschild radius $r_s$ will be of order $M^{1/(D-3)}\sim 
\TT^{1/(D-3)} L^{(D-k-1)/(D-3)}$.
For $k>2$ this growth is slower than $L$  for large $L$
and hence for sufficiently large $L$ the brane
is always outside its would be Schwarzschild radius. 
For codimension two branes we get $r_s\sim \TT^{1/(D-3)}\, L$ and hence whether or not
this is larger than $L$ depends on the tension of the brane.\footnote{The codimension two
branes we consider here have logarithmically varying scalar fields away from the brane.
So one might expect that the 
energy of a codimension two brane loop may have an additional factor of  
$\log L$ from the scalar field energy density and spoil the scaling 
argument given here\cite{1011.5120}.
We shall discuss in section \ref{sscaling} why this does not happen for the branes that
we shall study. \label{fo1}}
Finally for codimension one
brane we have $r_s\sim \TT^{1/(D-3)} L^{(D-2)/(D-3)}$ and for large $L$ this always exceeds
the size $L$ of the system. Thus in this case an asymptotic observer cannot study a large loop
of such branes.

This shows that the codimension two branes are somewhat special since 
our ability to study them
depends on the tension of the brane. 
String theory contains many
codimension two branes, including 
exotic branes with the property that as we transport a
state around such an
exotic brane, the state undergoes a non-trivial 
monodromy transformation associated with some duality
symmetry\cite{9302038,9408083,9707217,9712047,9712075,9712084,
9809039,1004.2521,1209.6056}.  The existence of these branes can be used to argue that
we must identify field configurations related by these duality transformations and hence
the corresponding symmetry is a gauge symmetry. On the other hand 
at weak coupling, the tension of most exotic branes
become large in the Planck scale and hence they are shielded by an event horizon and
become inaccessible to an asymptotic observer. Therefore the question arises as to in what sense
these branes exist and how we can argue that the monodromies associated with these branes
are gauge symmetries.

For BPS exotic branes we suggest a possible way out of this. 
While they may be behind the event
horizon at weak coupling, there is some corner of the moduli space where their tension
becomes small in the Planck units, since typically they are related to fundamental strings,
D-branes or other known objects via various duality symmetries, and the latter do become
light in some corner of the moduli space. Now it was argued in 
\cite{2502.07883,2503.00601,2506.13876} that irrespective of the
asymptotic values of the moduli, string theory contains states for which in an arbitrary large
region in space-time the moduli take any desired value, and furthermore, this region is not
hidden from the asymptotic observer by an event horizon, thus avoiding the
issues discussed in \cite{2501.17697}. So one could in principle have a
state in which in a large region of space-time the exotic brane is light and is not hidden by an
event horizon. Since this region is accessible to an asymptotic observer, such an observer
can study the exotic brane and measure the monodromy associated with these branes, thereby
establishing that these symmetries are indeed gauged. 
We discuss this for various types of codimension two branes in
sections \ref{s2}, \ref{s3} and \ref{s4}, and argue that for all exotic BPS branes we can find
appropriate regions in the moduli space where their tension becomes small in Planck units,
and hence these brane loops can be studied by creating a region in space-time where the
moduli take the desired values.

In section \ref{sscaling} we address the problem described in footnote \ref{fo1}, -- the
possibility of additional logarithmic factors in the energy of codimension two brane
loops due to logarithmically varying scalar fields produced by the brane.
We show that 
the scaling argument, that shows that the energy of the brane loop of size $L$ should grow as
$L^{D-3}$, relies on the hidden assumption that
the energy of the brane does not depend on the microscopic cut-off given by the thickness of the
brane. We then  show how the variation of the dilaton removes the dependence on the
microscopic cut-off for the
BPS branes of the type analyzed in sections  \ref{s2}, \ref{s3} and \ref{s4}, and reproduces the
result based on the scaling argument.

As already mentioned, monodromies around codimension two branes form part of
discrete gauge symmetries of the theory. We discuss this in detail
in section \ref{sudual} and use this to formulate a precise test of discrete gauge symmetries in
string theory. This extends to the cases when the discrete symmetry is spontaneously broken,
includng the U-duality symmetries of the theory, all of which are spontaneously broken
at a generic point in the moduli space of the theory.

The last point may seem somewhat surprising, since it was argued in 
\cite{1810.05337,1810.05338} that in anti de Sitter space-time, spontaneously broken
gauge symmetries have no observable signature. Indeed type IIB string theory
on $AdS_5\times S^5$ has SL(2,$\ZZZ)$
S-duality symmetry that can be identified as the S-duality symmetry of the
holographically dual N=4 supersymmetric Yang-Mills theory.  However, for generic values of the
complex coupling, the S-duality symmetry of the Yang-Mills theory is explicitly broken and has no
direct observational signature, except possibly some BPS states whose existence is protected
by supersymmetry\cite{9402032}. 
In particular the status of S-duality is on the same footing as the $\phi\to -\phi$ symmetry in a 
scalar field theory with $\lambda\phi^3$ interaction. The transformation relates theories at
$\lambda$ and $-\lambda$ but is not a symmetry unless $\lambda$ vanishes.
We discuss in section \ref{sads} why the situation in AdS is
different from that in flat space-time where we do
have observable signature of the duality symmetry even for 
generic values of the asymptotic moduli.

For non-BPS exotic branes, some of which have been conjectured to exist in string theory, the
situation is more complicated, since we do not have {\it a priori} knowledge of their tension at
any point in the moduli space. We discuss this in section \ref{s6}.

In sections \ref{s5} and \ref{sflat} we take a different approach 
to studying low codimension branes.
Instead of studying them via finite energy configurations in the theory of interest, we
start with a different string theory, 
and then realize both the desired string theory and the low
codimension branes in this theory as a configuration in the parent string theory. 

In section \ref{s5} we
exploit the fact that in many
string compactifications, various low codimension branes are already present as part of
compactfication. These branes extend along the non-compact part of the space-time but 
are localized along one or more compact internal directions. One class of 
examples of this type is
F-theory compactifications where $(p,q)$ seven branes, localized along two of the compact
directions and extended along other space-time directions are present in the 
system\cite{9602022,9602114,9603161}.
Another example is provided by
$E_8\times E_8$ heterotic string theory
whose strong coupling limit is M-theory `compactified' on a 
line segment with end of the world 9-branes at the two ends of the segment, each carrying one
set of $E_8$ gauge fields\cite{9510209}. A third example involves 
type I$'$ theory\cite{9510169} where
we have type IIA orientifold 8-planes sitting at the two ends of an interval, with D8-branes
parallel to the orientifold planes placed at various points along the interval and the regions
between the D8-branes and orientifold planes are
described by massive type IIA supergravity
discovered by Romans\cite{romans}.
By creating large regions in space-time where the size 
of the compact directions transverse and tangential
to the branes becomes large, we can create a configuration where we have isolated flat
branes that can be studied by an asymptotic observer.

In section \ref{sflat} we consider a hypothetical flat codimension one
brane, carrying positive energy density,
that separates two Minkowski vacua A and B of string theory. 
Since the brane has infinite energy, it is not a
state in theory A or theory B, but it is an allowed background in string theory as long as such a
brane exists. We then
explore to what extent an observer in
A can explore the properties of B and vice versa. We find that even in this case, an observer
remaining in A cannot study the properties of B to arbitrary precision, since there is a limit to
how long an apparatus can remain in B in an inertial frame  and still send back signal to the
observer in A. It is however possible for an observer to
spend infinite amount of time in A, finish all the desired measurements and then cross over to
B and spend infinite amount of time there, performing all the desired measurements in B. Such
an observer will have complete information on the observables of A and B,\footnote{In this paper,
by observables of a theory we shall mean all the S-matrix elements of that theory.}
but the price we
pay is that the local environment of the incoming observer will 
differ from that of an outgoing observer.

\sectiono{SL(2,$\ZZZ$) multiplet of  strings in four dimensions} \label{s2}

We begin our discussion with the SL(2,$\ZZZ$) family of strings discussed in \cite{9302038}.
Heterotic string theory on $T^6\times \RRR^{3,1}$ is invariant under an SL(2,$\ZZZ$)
symmetry:
\be \label{etautrs}
\tau \to \tau'= {p\tau+q\over r\tau +s }, \qquad  \pmatrix{p & q\cr r & s} \in
{\rm SL(2},\ZZZ)\, ,
\ee
where
\be
\tau = a_0 + i \, g_4^{-2}\equiv \tau_1 + i\tau_2\, ,
\ee
Here $g_4$ is a 
suitably normalized four dimensional string coupling, related to the asymptotic value
of $e^\phi$ where $\phi$ is the dilaton and $a_0$ is  the asymptotic value of the axion,
obtained by dualizing the 2-form field $B_{\mu\nu}$. The
fundamental string, which we shall call the (1,0) string, has tension $1/(2\pi \ell_s^2)$
where $\ell_s=\sqrt{\alpha'}$ is the string length scale. Since in $D$ space-time dimensions 
the Planck length $\ell_p$ is proportional to $\ell_s g_D^{2/(D-2)}$, $g_D$ being the $D$-dimensional
string coupling, we have 
$\ell_p\propto \ell_s g_4$ in $D=4$. Therefore, if we work in units where the 
Planck length $\ell_p$ is set equal to one, then the tension $T_{1,0}$ of the fundamental string would
be given by,
\be \label{etension}
T_{1,0} = C \, g_4^2 = C \tau_2^{-1}\, ,
\ee
where $C$ is a numerical constant.
Throughout this paper we shall work in  units where the Planck length is set equal to 
one since under a duality transformation the Planck length remains
unchanged.
If we travel along a closed path around the 
fundamental string, $\tau$ gets transformed by the
SL(2,$\ZZZ$) transformation
\be \label{emonodromyone}
g_{1,0} = \pmatrix{1 & 1\cr 0 & 1}\, .
\ee
We call $g_{1,0}$ the monodromy associated with the fundamental string.

We can combine the SL(2,$\ZZZ$) symmetry with the existence of the fundamental string
to predict the existence of exotic strings\cite{9302038}. 
For this consider the effect of SL(2,$\ZZZ$) transformation
on a fundamental string. 
If there is a fundamental string at coupling $\tau'=(p\tau+q)/(r\tau+s)$, then it should be
related to a new string at coupling $\tau$, which we shall call the $(s,r)$ string, with tension 
\be 
T_{s,r} = C\tau_2'{}^{-1}= C {|r\tau+s|^2\over \tau_2} \, .
\ee 
If we travel along a closed path around the 
$(s,r)$ string, $\tau$ gets transformed by the
SL(2,$\ZZZ$) transformation
\be\label{emonodromy}
g_{s,r}\equiv \pmatrix{p & q\cr r & s}^{-1} \pmatrix{1 & 1\cr 0 & 1} \pmatrix{p & q\cr r & s}
= \pmatrix{1+rs & s^2 \cr -r^2 & 1-rs}\, .
 \ee
The existence of the monodromy implies that we must identify field configurations related by
the duality transformation \refb{emonodromy} and hence this must be a discrete gauge symmetry.
For a generic value of $\tau$ this is spontaneously broken since $\tau$ is not invariant under
this symmetry. 

We shall now review how we could measure this monodromy experimentally.
The  SL(2,$\ZZZ$) transformation \refb{etautrs} acts on the quantized electric and magnetic charge
vectors $(Q,P)$, for appropriate choice of signs of these charges, as
\be\label{echargemon}
\pmatrix{Q\cr P}\to \pmatrix{Q'\cr P'}=g \pmatrix{Q\cr P}\, .
\ee
Now consider taking a particle carrying charges $(Q,P)$ and 
adiabatically transporting it along a
closed curve around the string.
Since $Q,P$ take values on a fixed lattice, they cannot change
continuously and remain fixed during this transport even though
$\tau$ changes continuously and returns to a value that is related to its original
value by 
the SL(2,$\ZZZ)$ transformation \refb{emonodromy}. To see how the
transported particle appears to
an observer who has not gone around the string, we need to
bring $\tau$ back to the original value by the action of the inverse of
\refb{emonodromy}.
This transforms the charges according to \refb{echargemon} with $g$ given by
the inverse of \refb{emonodromy}.
This can be measured by starting
with two particles carrying identical charges, taking one of them around the string and
then comparing the charges of the two particles.

In the above discussion we have considered the string to be infinitely long.
Such a configuration has infinite mass and changes the asymptotic structure of space-time.
So if we want to work in asymptotically flat space-time, we need to consider a loop of
this string. If we take the loop to be sufficiently large then locally it will look like
a straight string and any experiment we would like to perform on an infinitely long 
string with finite resources can also be performed on such macroscopic strings, since in any
case with finite resources we cannot explore properties of an infinitely long string. In particular
the tension \refb{etension} and the monodromy property \refb{emonodromy} can be measured
using experiments confined near a segment of the macroscopic string.
This however relies on one assumption -- that the string is accessible to the
asymptotic observer and is not hidden behind a black hole horizon. To check this we note that
a macroscopic $(s,r)$ string of length $L$ has mass of order
\be
L\, T_{s,r} = C\, L\, {|r\tau+s|^2 \over \tau_2}\, 
\ee
in Planck units.
Hence its Schwarzschild radius $\propto 2m$ is of order
\be
R_{s,r}\equiv  L\, \, {|r\tau+s|^2 \over \tau_2}\, ,
\ee
The reason for stating the mass and the Schwarzschild radius
as an order of magnitude estimate
instead of exact numbers is 
due to the fact that a loop of string is not spherically symmetric and it acts as a source of
the axion-dilaton field besides gravity. Hence we need to take into account the energy density
in the axion-dilaton field and the effect of lack of spherical symmetry on the solution.
The exact expressions can in principle be calculated by solving the full set of field equations
in the presence of a source corresponding to a circular loop of string.
Nevertheless it is clear that if $R_{s,r}<<L$
then the string will be accessible to an asymptotic observer, while if $R_{s,r}>>L$ then
the string will be behind the event horizon and will not be accessible to the asymptotic
observer.\footnote{One could worry whether the energy contained in the fields can make the
total energy and hence $R_{s,r}$
proportional to $L\ln L$\cite{1011.5120}. We shall address this in section \ref{sscaling}.}
In particular, for the fundamental string $(s,r)=(1,0)$,
and $R_{1,0}=2\, C\, L\, \tau_2^{-1}$ is much smaller than $L$ in the weak coupling limit
$\tau_2\to\infty$
and is accessible to the asymptotic observer. However, $R_{1,0}$ is much larger that $L$ in the
strong coupling limit and hence the string loop will be behind its event horizon and will not be
accessible to the asymptotic observer. More generally, for the $(s,r)$ string, we have $R_{s,r}
<< L$ in the neighborhood of the point $\tau=-s/r$, but for finite $\tau$,
we have $R_{s,r}\sim L$. If the asymptotic
coupling is weak, i.e. $\tau_2>>1$, then all the $(s,r)$ strings with $r\ne 0$ will have
$R_{s,r}>>L$ and will be behind the horizon.

This raises the question: in what sense the general $(s,r)$ strings exist in an asymptotically
weakly coupled theory? To this end, note that it was shown in 
\cite{2502.07883,2503.00601,2506.13876} that in an asymptotically flat
string theory with a moduli space of vacua, it is possible to create an arbitrarily large space-time region
in which the moduli can take any desired value and the curvature, other field strengths and scalar
field gradients can be made as small as one likes. 
In particular, if we take a purely electrically charged near extremal
black hole in the heterotic string theory on $T^6$,
carrying both, momentum and winding charges, along some compact directions, 
its near horizon geometry has weak string coupling where the fundamental string has
low tension. It follows from SL(2,$\ZZZ$) invariance that a
dyonic near extremal black hole carrying electric and magnetic charges of the form
$(sQ,-rQ)$, will have  $\tau \simeq -s/r$ in its
near horizon region.
By scaling up the mass and charge of such a black
hole,
one can create an arbitrarily large space-time region where
$\tau \simeq -s/r$ so that
the $(s,r)$ string is light in this region, and perform experiments on the $(s,r)$ string to study its 
properties.\footnote{One could object that to an observer in the near horizon region
the $(s,r)$ string will appear
as a fundamental string at weak coupling and hence this is not a test of existence of a new
kind of string. However asymptotic observers could send charged probe particles to the
experimentalists in the near horizon region of the black hole in advance to establish what
is electric and what is magnetic charge. After that there will be no ambiguity in
distinguishing $(s,r)$ strings from fundamental strings.}

In the above discussion we have not discussed in detail the mechanism for producing macroscopic
string loops. We take the viewpoint that as long as we are considering a state in the theory it can
always be produced, {\it e.g.} in a high energy collision or the decay of a heavy black hole 
such states may be produced with
small but non-zero probability. For this the $(s,r)$ strings 
are on the same footing as the
fundamental strings since if we can produce macroscopic fundamental string in 
weakly coupled string
theory we can produce $(s,r)$ strings by the 
S-dual process in the region of space-time where the
$(s,r)$ strings are light.

\sectiono{SL(2,$\ZZZ$) multiplet of seven branes in type IIB  theory} \label{s3}

The case of $(s,r)$ seven branes in type IIB string theory is very similar. The
complex coupling is given by
\be 
\tau = a + {i\over g_s} \, ,
\ee
where $a$ is the asymptotic value of the RR scalar field and $g_s$ is the ten dimensional 
string
coupling. The tension  of the D7-brane, also denoted as the (1,0) brane, is given by\be
\TT_{1,0} \propto g_s^{-1} / \ell_s^{8}\, .
\ee
In ten dimensions, $\ell_p = g_s^{1/4}\ell_s$. Hence in the Planck units $\ell_s\sim g_s^{-1/4}$
and we get
\be
\TT_{1,0}
=\CC\, 
g_s^{-1} g_s^2 = \CC\, g_s\, ,
\ee
for some numerical constant $\CC$. In terms of $\tau$, this takes the form:
\be
\TT_{1,0} = \CC \tau_2^{-1}\, .
\ee
The S-duality transformation acts on $\tau$ in the same way
as \refb{etautrs} and produces an $(s,r)$ 7-brane with tension:
\be
\TT_{s,r} = \CC  \, {|r\tau+s|^2\over \tau_2}\, .
\ee
A closed loop surrounding a D7-brane has monodromy identical to the one given in 
\refb{emonodromyone}. Using S-duality invariance of the theory
one can conclude that an $(s,r)$ 7-brane will produce a monodromy \refb{emonodromy}.

Since flat 7-branes in ten space-time dimensions have infinite mass and changes the
asymptotic structure of space-time, they are by themselves not states of the system. However
we can construct finite energy states from macroscopic loops of seven branes, where at some
given instant in time, the
spatial part of the seven brane world-volume lie along a compact subspace of $\RRR^9$.
In particular, we can consider an $(s,r)$ seven brane lying along a seven dimensional sphere of radius
$L$. Up to a numerical factor, its mass will be given by $L^7 \TT_{s,r}$ and in 9+1 dimension,
its Schwarzschild
radius $R_{s,r}$ will be proportional to the $1/7$-th power of its mass. This gives
\be
R_{s,r} = \KK\, L\, |r\, \tau+s|^{2/7}\, \tau_2^{-1/7}\, ,
\ee
where $\KK$ is a numerical constant. From this we see that the spherical $(s,r)$ brane have
Schwarzscild radius small compared to the size $L$ of the system
only near the region $\tau\simeq -s/r$. In particular in the weak coupling limit $\tau_2\to\infty$,
all the $(s,r)$ branes other than the D7-brane have Schwarzschild radius much larger than the
size $L$ of the system and hence such branes cannot be studied by an asymptotic observer.

The resolution of this issue is also similar to the one used for the four dimensional string theory,
namely we can create large regions of space-time where the modulus $\tau$ take values close to
$-s/r$, and in that region we can have macroscopic loops of $(s,r)$ seven branes and study their
properties. The only new ingredient is that the ten dimensional type IIB string theory does not
have any charged black holes that can be used to produce such regions. However we can use
thick domain walls described in \cite{2506.13876} for this purpose.

\sectiono{Other codimension two BPS branes}  \label{s4}

Ref.~\cite{1209.6056} listed various other BPS exotic branes in string theory. The
construction of these branes follows paths similar to the one described in the last
two sections, namely in a given compactified string theory, one starts with some known
codimension two branes constructed from
fundamental strings, D-branes, NS 5-branes or Kaluza-Klein (KK) monopoles
and then uses U-duality transformation to 
construct other codimension two branes. 
We shall call the original branes the primary branes and the new branes obtained via
U-duality transformation the secondary branes.

Now, each of the primary branes have tension small
compared to the Planck scale in some region of the moduli space. We have already seen that for
D-branes and fundamental strings the tension becomes small when the string coupling becomes
small. For NS 5-branes we can make the tension small in Planck units by taking
the compact directions transverse to the NS 5-brane to 
have large size since in this limit the Newton's constant
in the compactified theory becomes small and hence the Planck mass becomes large
as measured in string scale. For KK monopoles we can make the tension small by taking
the circle associated with the KK monopole to have small size. 
Then by U-duality invariance of the BPS mass formula, measured in the Planck units,
it follows that 
if the primary brane becomes light in some region $A$ of the moduli space,
then the secondary brane, obtained from this 
primary brane by some U-duality transformation $g$, becomes light in the
region $B=g(A)$ of the moduli space, where $g(A)$ denotes 
$g$ transformation of the region $A$. 
We can then create large regions in space-time where the moduli
takes value in region $B$, and in this region a macroscopic loop of the secondary brane
will have its Schwarzschild radius small compared to its size. Hence an asymptotic observer
will be able to study these branes even if the asymptotic coupling is weak and these branes
do not exist in the asymptotic region.

One common feature of the construction of exotic branes described above is that while in a
given `theory' characterized by fixed asymptotic values of the moduli we can produce all BPS
exotic branes by producing appropriate values of the moduli fields in different regions of
space-time, we cannot necessarily bring these branes together. This avoids some apparent puzzles
that would otherwise be present. For example in type IIB string theory, the total monodromy around
a (2,1) and (0,1) seven brane is given by:
\be
 \pmatrix{3 & 4 \cr -1 & -1}\pmatrix{1 & 0 \cr -1 & 1}
= \pmatrix{-1 & 4 \cr 0 & -1}\, .
\ee
This is the monodromy around an orientifold seven 
plane\cite{9605150}. Thus if we could bring a macroscopic
loop of (2,1) and (0,1) seven branes together than they should produce a macroscopic loop
of orientifold
seven plane. This however leads to a puzzle since the $(2,1)$ and (0,1) seven branes have
zero modes corresponding to translation in 
transverse directions while an orientifold seven plane
does not have these modes. Indeed, an orientifold plane has negative tension and a 
dynamical loop of orientifold plane will render the theory inconsistent.
However if (depending on the ambient values of the moduli)
either  the
(2,1) seven brane or the (0,1) seven brane is hidden behind an event horizon,  then the
whole system will be behind an event horizon and there is no contradiction.

\sectiono{Scaling of macroscopic codimension two brane loops} \label{sscaling}

\def\figstringloop{

\def\JPicScale{0.4}
\ifx\JPicScale\undefined\def\JPicScale{1}\fi
\unitlength \JPicScale mm
\begin{picture}(130,80)(0,0)
\linethickness{0.3mm}
\put(103.96,39.75){\line(0,1){0.5}}
\multiput(103.95,40.75)(0.01,-0.5){1}{\line(0,-1){0.5}}
\multiput(103.94,41.25)(0.01,-0.5){1}{\line(0,-1){0.5}}
\multiput(103.91,41.74)(0.02,-0.5){1}{\line(0,-1){0.5}}
\multiput(103.89,42.24)(0.03,-0.5){1}{\line(0,-1){0.5}}
\multiput(103.85,42.74)(0.04,-0.5){1}{\line(0,-1){0.5}}
\multiput(103.81,43.24)(0.04,-0.5){1}{\line(0,-1){0.5}}
\multiput(103.75,43.73)(0.05,-0.5){1}{\line(0,-1){0.5}}
\multiput(103.7,44.23)(0.06,-0.5){1}{\line(0,-1){0.5}}
\multiput(103.63,44.72)(0.07,-0.49){1}{\line(0,-1){0.49}}
\multiput(103.56,45.21)(0.07,-0.49){1}{\line(0,-1){0.49}}
\multiput(103.48,45.71)(0.08,-0.49){1}{\line(0,-1){0.49}}
\multiput(103.39,46.2)(0.09,-0.49){1}{\line(0,-1){0.49}}
\multiput(103.29,46.69)(0.09,-0.49){1}{\line(0,-1){0.49}}
\multiput(103.19,47.17)(0.1,-0.49){1}{\line(0,-1){0.49}}
\multiput(103.08,47.66)(0.11,-0.49){1}{\line(0,-1){0.49}}
\multiput(102.97,48.15)(0.12,-0.48){1}{\line(0,-1){0.48}}
\multiput(102.85,48.63)(0.12,-0.48){1}{\line(0,-1){0.48}}
\multiput(102.71,49.11)(0.13,-0.48){1}{\line(0,-1){0.48}}
\multiput(102.58,49.59)(0.14,-0.48){1}{\line(0,-1){0.48}}
\multiput(102.43,50.07)(0.14,-0.48){1}{\line(0,-1){0.48}}
\multiput(102.28,50.54)(0.15,-0.48){1}{\line(0,-1){0.48}}
\multiput(102.12,51.01)(0.16,-0.47){1}{\line(0,-1){0.47}}
\multiput(101.96,51.48)(0.17,-0.47){1}{\line(0,-1){0.47}}
\multiput(101.79,51.95)(0.17,-0.47){1}{\line(0,-1){0.47}}
\multiput(101.61,52.42)(0.18,-0.47){1}{\line(0,-1){0.47}}
\multiput(101.42,52.88)(0.09,-0.23){2}{\line(0,-1){0.23}}
\multiput(101.23,53.34)(0.1,-0.23){2}{\line(0,-1){0.23}}
\multiput(101.03,53.8)(0.1,-0.23){2}{\line(0,-1){0.23}}
\multiput(100.82,54.25)(0.1,-0.23){2}{\line(0,-1){0.23}}
\multiput(100.61,54.7)(0.11,-0.23){2}{\line(0,-1){0.23}}
\multiput(100.39,55.15)(0.11,-0.22){2}{\line(0,-1){0.22}}
\multiput(100.17,55.59)(0.11,-0.22){2}{\line(0,-1){0.22}}
\multiput(99.94,56.04)(0.12,-0.22){2}{\line(0,-1){0.22}}
\multiput(99.7,56.47)(0.12,-0.22){2}{\line(0,-1){0.22}}
\multiput(99.45,56.91)(0.12,-0.22){2}{\line(0,-1){0.22}}
\multiput(99.2,57.34)(0.13,-0.22){2}{\line(0,-1){0.22}}
\multiput(98.94,57.77)(0.13,-0.21){2}{\line(0,-1){0.21}}
\multiput(98.68,58.19)(0.13,-0.21){2}{\line(0,-1){0.21}}
\multiput(98.41,58.61)(0.14,-0.21){2}{\line(0,-1){0.21}}
\multiput(98.13,59.02)(0.14,-0.21){2}{\line(0,-1){0.21}}
\multiput(97.85,59.43)(0.14,-0.21){2}{\line(0,-1){0.21}}
\multiput(97.56,59.84)(0.14,-0.2){2}{\line(0,-1){0.2}}
\multiput(97.27,60.24)(0.15,-0.2){2}{\line(0,-1){0.2}}
\multiput(96.97,60.64)(0.1,-0.13){3}{\line(0,-1){0.13}}
\multiput(96.66,61.03)(0.1,-0.13){3}{\line(0,-1){0.13}}
\multiput(96.35,61.42)(0.1,-0.13){3}{\line(0,-1){0.13}}
\multiput(96.03,61.81)(0.11,-0.13){3}{\line(0,-1){0.13}}
\multiput(95.71,62.19)(0.11,-0.13){3}{\line(0,-1){0.13}}
\multiput(95.38,62.56)(0.11,-0.13){3}{\line(0,-1){0.13}}
\multiput(95.05,62.93)(0.11,-0.12){3}{\line(0,-1){0.12}}
\multiput(94.71,63.3)(0.11,-0.12){3}{\line(0,-1){0.12}}
\multiput(94.36,63.66)(0.11,-0.12){3}{\line(0,-1){0.12}}
\multiput(94.01,64.01)(0.12,-0.12){3}{\line(0,-1){0.12}}
\multiput(93.66,64.36)(0.12,-0.12){3}{\line(1,0){0.12}}
\multiput(93.3,64.71)(0.12,-0.11){3}{\line(1,0){0.12}}
\multiput(92.93,65.05)(0.12,-0.11){3}{\line(1,0){0.12}}
\multiput(92.56,65.38)(0.12,-0.11){3}{\line(1,0){0.12}}
\multiput(92.19,65.71)(0.13,-0.11){3}{\line(1,0){0.13}}
\multiput(91.81,66.03)(0.13,-0.11){3}{\line(1,0){0.13}}
\multiput(91.42,66.35)(0.13,-0.11){3}{\line(1,0){0.13}}
\multiput(91.03,66.66)(0.13,-0.1){3}{\line(1,0){0.13}}
\multiput(90.64,66.97)(0.13,-0.1){3}{\line(1,0){0.13}}
\multiput(90.24,67.27)(0.13,-0.1){3}{\line(1,0){0.13}}
\multiput(89.84,67.56)(0.2,-0.15){2}{\line(1,0){0.2}}
\multiput(89.43,67.85)(0.2,-0.14){2}{\line(1,0){0.2}}
\multiput(89.02,68.13)(0.21,-0.14){2}{\line(1,0){0.21}}
\multiput(88.61,68.41)(0.21,-0.14){2}{\line(1,0){0.21}}
\multiput(88.19,68.68)(0.21,-0.14){2}{\line(1,0){0.21}}
\multiput(87.77,68.94)(0.21,-0.13){2}{\line(1,0){0.21}}
\multiput(87.34,69.2)(0.21,-0.13){2}{\line(1,0){0.21}}
\multiput(86.91,69.45)(0.22,-0.13){2}{\line(1,0){0.22}}
\multiput(86.47,69.7)(0.22,-0.12){2}{\line(1,0){0.22}}
\multiput(86.04,69.94)(0.22,-0.12){2}{\line(1,0){0.22}}
\multiput(85.59,70.17)(0.22,-0.12){2}{\line(1,0){0.22}}
\multiput(85.15,70.39)(0.22,-0.11){2}{\line(1,0){0.22}}
\multiput(84.7,70.61)(0.22,-0.11){2}{\line(1,0){0.22}}
\multiput(84.25,70.82)(0.23,-0.11){2}{\line(1,0){0.23}}
\multiput(83.8,71.03)(0.23,-0.1){2}{\line(1,0){0.23}}
\multiput(83.34,71.23)(0.23,-0.1){2}{\line(1,0){0.23}}
\multiput(82.88,71.42)(0.23,-0.1){2}{\line(1,0){0.23}}
\multiput(82.42,71.61)(0.23,-0.09){2}{\line(1,0){0.23}}
\multiput(81.95,71.79)(0.47,-0.18){1}{\line(1,0){0.47}}
\multiput(81.48,71.96)(0.47,-0.17){1}{\line(1,0){0.47}}
\multiput(81.01,72.12)(0.47,-0.17){1}{\line(1,0){0.47}}
\multiput(80.54,72.28)(0.47,-0.16){1}{\line(1,0){0.47}}
\multiput(80.07,72.43)(0.48,-0.15){1}{\line(1,0){0.48}}
\multiput(79.59,72.58)(0.48,-0.14){1}{\line(1,0){0.48}}
\multiput(79.11,72.71)(0.48,-0.14){1}{\line(1,0){0.48}}
\multiput(78.63,72.85)(0.48,-0.13){1}{\line(1,0){0.48}}
\multiput(78.15,72.97)(0.48,-0.12){1}{\line(1,0){0.48}}
\multiput(77.66,73.08)(0.48,-0.12){1}{\line(1,0){0.48}}
\multiput(77.17,73.19)(0.49,-0.11){1}{\line(1,0){0.49}}
\multiput(76.69,73.29)(0.49,-0.1){1}{\line(1,0){0.49}}
\multiput(76.2,73.39)(0.49,-0.09){1}{\line(1,0){0.49}}
\multiput(75.71,73.48)(0.49,-0.09){1}{\line(1,0){0.49}}
\multiput(75.21,73.56)(0.49,-0.08){1}{\line(1,0){0.49}}
\multiput(74.72,73.63)(0.49,-0.07){1}{\line(1,0){0.49}}
\multiput(74.23,73.7)(0.49,-0.07){1}{\line(1,0){0.49}}
\multiput(73.73,73.75)(0.5,-0.06){1}{\line(1,0){0.5}}
\multiput(73.24,73.81)(0.5,-0.05){1}{\line(1,0){0.5}}
\multiput(72.74,73.85)(0.5,-0.04){1}{\line(1,0){0.5}}
\multiput(72.24,73.89)(0.5,-0.04){1}{\line(1,0){0.5}}
\multiput(71.74,73.91)(0.5,-0.03){1}{\line(1,0){0.5}}
\multiput(71.25,73.94)(0.5,-0.02){1}{\line(1,0){0.5}}
\multiput(70.75,73.95)(0.5,-0.01){1}{\line(1,0){0.5}}
\multiput(70.25,73.96)(0.5,-0.01){1}{\line(1,0){0.5}}
\put(69.75,73.96){\line(1,0){0.5}}
\multiput(69.25,73.95)(0.5,0.01){1}{\line(1,0){0.5}}
\multiput(68.75,73.94)(0.5,0.01){1}{\line(1,0){0.5}}
\multiput(68.26,73.91)(0.5,0.02){1}{\line(1,0){0.5}}
\multiput(67.76,73.89)(0.5,0.03){1}{\line(1,0){0.5}}
\multiput(67.26,73.85)(0.5,0.04){1}{\line(1,0){0.5}}
\multiput(66.76,73.81)(0.5,0.04){1}{\line(1,0){0.5}}
\multiput(66.27,73.75)(0.5,0.05){1}{\line(1,0){0.5}}
\multiput(65.77,73.7)(0.5,0.06){1}{\line(1,0){0.5}}
\multiput(65.28,73.63)(0.49,0.07){1}{\line(1,0){0.49}}
\multiput(64.79,73.56)(0.49,0.07){1}{\line(1,0){0.49}}
\multiput(64.29,73.48)(0.49,0.08){1}{\line(1,0){0.49}}
\multiput(63.8,73.39)(0.49,0.09){1}{\line(1,0){0.49}}
\multiput(63.31,73.29)(0.49,0.09){1}{\line(1,0){0.49}}
\multiput(62.83,73.19)(0.49,0.1){1}{\line(1,0){0.49}}
\multiput(62.34,73.08)(0.49,0.11){1}{\line(1,0){0.49}}
\multiput(61.85,72.97)(0.48,0.12){1}{\line(1,0){0.48}}
\multiput(61.37,72.85)(0.48,0.12){1}{\line(1,0){0.48}}
\multiput(60.89,72.71)(0.48,0.13){1}{\line(1,0){0.48}}
\multiput(60.41,72.58)(0.48,0.14){1}{\line(1,0){0.48}}
\multiput(59.93,72.43)(0.48,0.14){1}{\line(1,0){0.48}}
\multiput(59.46,72.28)(0.48,0.15){1}{\line(1,0){0.48}}
\multiput(58.99,72.12)(0.47,0.16){1}{\line(1,0){0.47}}
\multiput(58.52,71.96)(0.47,0.17){1}{\line(1,0){0.47}}
\multiput(58.05,71.79)(0.47,0.17){1}{\line(1,0){0.47}}
\multiput(57.58,71.61)(0.47,0.18){1}{\line(1,0){0.47}}
\multiput(57.12,71.42)(0.23,0.09){2}{\line(1,0){0.23}}
\multiput(56.66,71.23)(0.23,0.1){2}{\line(1,0){0.23}}
\multiput(56.2,71.03)(0.23,0.1){2}{\line(1,0){0.23}}
\multiput(55.75,70.82)(0.23,0.1){2}{\line(1,0){0.23}}
\multiput(55.3,70.61)(0.23,0.11){2}{\line(1,0){0.23}}
\multiput(54.85,70.39)(0.22,0.11){2}{\line(1,0){0.22}}
\multiput(54.41,70.17)(0.22,0.11){2}{\line(1,0){0.22}}
\multiput(53.96,69.94)(0.22,0.12){2}{\line(1,0){0.22}}
\multiput(53.53,69.7)(0.22,0.12){2}{\line(1,0){0.22}}
\multiput(53.09,69.45)(0.22,0.12){2}{\line(1,0){0.22}}
\multiput(52.66,69.2)(0.22,0.13){2}{\line(1,0){0.22}}
\multiput(52.23,68.94)(0.21,0.13){2}{\line(1,0){0.21}}
\multiput(51.81,68.68)(0.21,0.13){2}{\line(1,0){0.21}}
\multiput(51.39,68.41)(0.21,0.14){2}{\line(1,0){0.21}}
\multiput(50.98,68.13)(0.21,0.14){2}{\line(1,0){0.21}}
\multiput(50.57,67.85)(0.21,0.14){2}{\line(1,0){0.21}}
\multiput(50.16,67.56)(0.2,0.14){2}{\line(1,0){0.2}}
\multiput(49.76,67.27)(0.2,0.15){2}{\line(1,0){0.2}}
\multiput(49.36,66.97)(0.13,0.1){3}{\line(1,0){0.13}}
\multiput(48.97,66.66)(0.13,0.1){3}{\line(1,0){0.13}}
\multiput(48.58,66.35)(0.13,0.1){3}{\line(1,0){0.13}}
\multiput(48.19,66.03)(0.13,0.11){3}{\line(1,0){0.13}}
\multiput(47.81,65.71)(0.13,0.11){3}{\line(1,0){0.13}}
\multiput(47.44,65.38)(0.13,0.11){3}{\line(1,0){0.13}}
\multiput(47.07,65.05)(0.12,0.11){3}{\line(1,0){0.12}}
\multiput(46.7,64.71)(0.12,0.11){3}{\line(1,0){0.12}}
\multiput(46.34,64.36)(0.12,0.11){3}{\line(1,0){0.12}}
\multiput(45.99,64.01)(0.12,0.12){3}{\line(1,0){0.12}}
\multiput(45.64,63.66)(0.12,0.12){3}{\line(0,1){0.12}}
\multiput(45.29,63.3)(0.11,0.12){3}{\line(0,1){0.12}}
\multiput(44.95,62.93)(0.11,0.12){3}{\line(0,1){0.12}}
\multiput(44.62,62.56)(0.11,0.12){3}{\line(0,1){0.12}}
\multiput(44.29,62.19)(0.11,0.13){3}{\line(0,1){0.13}}
\multiput(43.97,61.81)(0.11,0.13){3}{\line(0,1){0.13}}
\multiput(43.65,61.42)(0.11,0.13){3}{\line(0,1){0.13}}
\multiput(43.34,61.03)(0.1,0.13){3}{\line(0,1){0.13}}
\multiput(43.03,60.64)(0.1,0.13){3}{\line(0,1){0.13}}
\multiput(42.73,60.24)(0.1,0.13){3}{\line(0,1){0.13}}
\multiput(42.44,59.84)(0.15,0.2){2}{\line(0,1){0.2}}
\multiput(42.15,59.43)(0.14,0.2){2}{\line(0,1){0.2}}
\multiput(41.87,59.02)(0.14,0.21){2}{\line(0,1){0.21}}
\multiput(41.59,58.61)(0.14,0.21){2}{\line(0,1){0.21}}
\multiput(41.32,58.19)(0.14,0.21){2}{\line(0,1){0.21}}
\multiput(41.06,57.77)(0.13,0.21){2}{\line(0,1){0.21}}
\multiput(40.8,57.34)(0.13,0.21){2}{\line(0,1){0.21}}
\multiput(40.55,56.91)(0.13,0.22){2}{\line(0,1){0.22}}
\multiput(40.3,56.47)(0.12,0.22){2}{\line(0,1){0.22}}
\multiput(40.06,56.04)(0.12,0.22){2}{\line(0,1){0.22}}
\multiput(39.83,55.59)(0.12,0.22){2}{\line(0,1){0.22}}
\multiput(39.61,55.15)(0.11,0.22){2}{\line(0,1){0.22}}
\multiput(39.39,54.7)(0.11,0.22){2}{\line(0,1){0.22}}
\multiput(39.18,54.25)(0.11,0.23){2}{\line(0,1){0.23}}
\multiput(38.97,53.8)(0.1,0.23){2}{\line(0,1){0.23}}
\multiput(38.77,53.34)(0.1,0.23){2}{\line(0,1){0.23}}
\multiput(38.58,52.88)(0.1,0.23){2}{\line(0,1){0.23}}
\multiput(38.39,52.42)(0.09,0.23){2}{\line(0,1){0.23}}
\multiput(38.21,51.95)(0.18,0.47){1}{\line(0,1){0.47}}
\multiput(38.04,51.48)(0.17,0.47){1}{\line(0,1){0.47}}
\multiput(37.88,51.01)(0.17,0.47){1}{\line(0,1){0.47}}
\multiput(37.72,50.54)(0.16,0.47){1}{\line(0,1){0.47}}
\multiput(37.57,50.07)(0.15,0.48){1}{\line(0,1){0.48}}
\multiput(37.42,49.59)(0.14,0.48){1}{\line(0,1){0.48}}
\multiput(37.29,49.11)(0.14,0.48){1}{\line(0,1){0.48}}
\multiput(37.15,48.63)(0.13,0.48){1}{\line(0,1){0.48}}
\multiput(37.03,48.15)(0.12,0.48){1}{\line(0,1){0.48}}
\multiput(36.92,47.66)(0.12,0.48){1}{\line(0,1){0.48}}
\multiput(36.81,47.17)(0.11,0.49){1}{\line(0,1){0.49}}
\multiput(36.71,46.69)(0.1,0.49){1}{\line(0,1){0.49}}
\multiput(36.61,46.2)(0.09,0.49){1}{\line(0,1){0.49}}
\multiput(36.52,45.71)(0.09,0.49){1}{\line(0,1){0.49}}
\multiput(36.44,45.21)(0.08,0.49){1}{\line(0,1){0.49}}
\multiput(36.37,44.72)(0.07,0.49){1}{\line(0,1){0.49}}
\multiput(36.3,44.23)(0.07,0.49){1}{\line(0,1){0.49}}
\multiput(36.25,43.73)(0.06,0.5){1}{\line(0,1){0.5}}
\multiput(36.19,43.24)(0.05,0.5){1}{\line(0,1){0.5}}
\multiput(36.15,42.74)(0.04,0.5){1}{\line(0,1){0.5}}
\multiput(36.11,42.24)(0.04,0.5){1}{\line(0,1){0.5}}
\multiput(36.09,41.74)(0.03,0.5){1}{\line(0,1){0.5}}
\multiput(36.06,41.25)(0.02,0.5){1}{\line(0,1){0.5}}
\multiput(36.05,40.75)(0.01,0.5){1}{\line(0,1){0.5}}
\multiput(36.04,40.25)(0.01,0.5){1}{\line(0,1){0.5}}
\put(36.04,39.75){\line(0,1){0.5}}
\multiput(36.04,39.75)(0.01,-0.5){1}{\line(0,-1){0.5}}
\multiput(36.05,39.25)(0.01,-0.5){1}{\line(0,-1){0.5}}
\multiput(36.06,38.75)(0.02,-0.5){1}{\line(0,-1){0.5}}
\multiput(36.09,38.26)(0.03,-0.5){1}{\line(0,-1){0.5}}
\multiput(36.11,37.76)(0.04,-0.5){1}{\line(0,-1){0.5}}
\multiput(36.15,37.26)(0.04,-0.5){1}{\line(0,-1){0.5}}
\multiput(36.19,36.76)(0.05,-0.5){1}{\line(0,-1){0.5}}
\multiput(36.25,36.27)(0.06,-0.5){1}{\line(0,-1){0.5}}
\multiput(36.3,35.77)(0.07,-0.49){1}{\line(0,-1){0.49}}
\multiput(36.37,35.28)(0.07,-0.49){1}{\line(0,-1){0.49}}
\multiput(36.44,34.79)(0.08,-0.49){1}{\line(0,-1){0.49}}
\multiput(36.52,34.29)(0.09,-0.49){1}{\line(0,-1){0.49}}
\multiput(36.61,33.8)(0.09,-0.49){1}{\line(0,-1){0.49}}
\multiput(36.71,33.31)(0.1,-0.49){1}{\line(0,-1){0.49}}
\multiput(36.81,32.83)(0.11,-0.49){1}{\line(0,-1){0.49}}
\multiput(36.92,32.34)(0.12,-0.48){1}{\line(0,-1){0.48}}
\multiput(37.03,31.85)(0.12,-0.48){1}{\line(0,-1){0.48}}
\multiput(37.15,31.37)(0.13,-0.48){1}{\line(0,-1){0.48}}
\multiput(37.29,30.89)(0.14,-0.48){1}{\line(0,-1){0.48}}
\multiput(37.42,30.41)(0.14,-0.48){1}{\line(0,-1){0.48}}
\multiput(37.57,29.93)(0.15,-0.48){1}{\line(0,-1){0.48}}
\multiput(37.72,29.46)(0.16,-0.47){1}{\line(0,-1){0.47}}
\multiput(37.88,28.99)(0.17,-0.47){1}{\line(0,-1){0.47}}
\multiput(38.04,28.52)(0.17,-0.47){1}{\line(0,-1){0.47}}
\multiput(38.21,28.05)(0.18,-0.47){1}{\line(0,-1){0.47}}
\multiput(38.39,27.58)(0.09,-0.23){2}{\line(0,-1){0.23}}
\multiput(38.58,27.12)(0.1,-0.23){2}{\line(0,-1){0.23}}
\multiput(38.77,26.66)(0.1,-0.23){2}{\line(0,-1){0.23}}
\multiput(38.97,26.2)(0.1,-0.23){2}{\line(0,-1){0.23}}
\multiput(39.18,25.75)(0.11,-0.23){2}{\line(0,-1){0.23}}
\multiput(39.39,25.3)(0.11,-0.22){2}{\line(0,-1){0.22}}
\multiput(39.61,24.85)(0.11,-0.22){2}{\line(0,-1){0.22}}
\multiput(39.83,24.41)(0.12,-0.22){2}{\line(0,-1){0.22}}
\multiput(40.06,23.96)(0.12,-0.22){2}{\line(0,-1){0.22}}
\multiput(40.3,23.53)(0.12,-0.22){2}{\line(0,-1){0.22}}
\multiput(40.55,23.09)(0.13,-0.22){2}{\line(0,-1){0.22}}
\multiput(40.8,22.66)(0.13,-0.21){2}{\line(0,-1){0.21}}
\multiput(41.06,22.23)(0.13,-0.21){2}{\line(0,-1){0.21}}
\multiput(41.32,21.81)(0.14,-0.21){2}{\line(0,-1){0.21}}
\multiput(41.59,21.39)(0.14,-0.21){2}{\line(0,-1){0.21}}
\multiput(41.87,20.98)(0.14,-0.21){2}{\line(0,-1){0.21}}
\multiput(42.15,20.57)(0.14,-0.2){2}{\line(0,-1){0.2}}
\multiput(42.44,20.16)(0.15,-0.2){2}{\line(0,-1){0.2}}
\multiput(42.73,19.76)(0.1,-0.13){3}{\line(0,-1){0.13}}
\multiput(43.03,19.36)(0.1,-0.13){3}{\line(0,-1){0.13}}
\multiput(43.34,18.97)(0.1,-0.13){3}{\line(0,-1){0.13}}
\multiput(43.65,18.58)(0.11,-0.13){3}{\line(0,-1){0.13}}
\multiput(43.97,18.19)(0.11,-0.13){3}{\line(0,-1){0.13}}
\multiput(44.29,17.81)(0.11,-0.13){3}{\line(0,-1){0.13}}
\multiput(44.62,17.44)(0.11,-0.12){3}{\line(0,-1){0.12}}
\multiput(44.95,17.07)(0.11,-0.12){3}{\line(0,-1){0.12}}
\multiput(45.29,16.7)(0.11,-0.12){3}{\line(0,-1){0.12}}
\multiput(45.64,16.34)(0.12,-0.12){3}{\line(0,-1){0.12}}
\multiput(45.99,15.99)(0.12,-0.12){3}{\line(1,0){0.12}}
\multiput(46.34,15.64)(0.12,-0.11){3}{\line(1,0){0.12}}
\multiput(46.7,15.29)(0.12,-0.11){3}{\line(1,0){0.12}}
\multiput(47.07,14.95)(0.12,-0.11){3}{\line(1,0){0.12}}
\multiput(47.44,14.62)(0.13,-0.11){3}{\line(1,0){0.13}}
\multiput(47.81,14.29)(0.13,-0.11){3}{\line(1,0){0.13}}
\multiput(48.19,13.97)(0.13,-0.11){3}{\line(1,0){0.13}}
\multiput(48.58,13.65)(0.13,-0.1){3}{\line(1,0){0.13}}
\multiput(48.97,13.34)(0.13,-0.1){3}{\line(1,0){0.13}}
\multiput(49.36,13.03)(0.13,-0.1){3}{\line(1,0){0.13}}
\multiput(49.76,12.73)(0.2,-0.15){2}{\line(1,0){0.2}}
\multiput(50.16,12.44)(0.2,-0.14){2}{\line(1,0){0.2}}
\multiput(50.57,12.15)(0.21,-0.14){2}{\line(1,0){0.21}}
\multiput(50.98,11.87)(0.21,-0.14){2}{\line(1,0){0.21}}
\multiput(51.39,11.59)(0.21,-0.14){2}{\line(1,0){0.21}}
\multiput(51.81,11.32)(0.21,-0.13){2}{\line(1,0){0.21}}
\multiput(52.23,11.06)(0.21,-0.13){2}{\line(1,0){0.21}}
\multiput(52.66,10.8)(0.22,-0.13){2}{\line(1,0){0.22}}
\multiput(53.09,10.55)(0.22,-0.12){2}{\line(1,0){0.22}}
\multiput(53.53,10.3)(0.22,-0.12){2}{\line(1,0){0.22}}
\multiput(53.96,10.06)(0.22,-0.12){2}{\line(1,0){0.22}}
\multiput(54.41,9.83)(0.22,-0.11){2}{\line(1,0){0.22}}
\multiput(54.85,9.61)(0.22,-0.11){2}{\line(1,0){0.22}}
\multiput(55.3,9.39)(0.23,-0.11){2}{\line(1,0){0.23}}
\multiput(55.75,9.18)(0.23,-0.1){2}{\line(1,0){0.23}}
\multiput(56.2,8.97)(0.23,-0.1){2}{\line(1,0){0.23}}
\multiput(56.66,8.77)(0.23,-0.1){2}{\line(1,0){0.23}}
\multiput(57.12,8.58)(0.23,-0.09){2}{\line(1,0){0.23}}
\multiput(57.58,8.39)(0.47,-0.18){1}{\line(1,0){0.47}}
\multiput(58.05,8.21)(0.47,-0.17){1}{\line(1,0){0.47}}
\multiput(58.52,8.04)(0.47,-0.17){1}{\line(1,0){0.47}}
\multiput(58.99,7.88)(0.47,-0.16){1}{\line(1,0){0.47}}
\multiput(59.46,7.72)(0.48,-0.15){1}{\line(1,0){0.48}}
\multiput(59.93,7.57)(0.48,-0.14){1}{\line(1,0){0.48}}
\multiput(60.41,7.42)(0.48,-0.14){1}{\line(1,0){0.48}}
\multiput(60.89,7.29)(0.48,-0.13){1}{\line(1,0){0.48}}
\multiput(61.37,7.15)(0.48,-0.12){1}{\line(1,0){0.48}}
\multiput(61.85,7.03)(0.48,-0.12){1}{\line(1,0){0.48}}
\multiput(62.34,6.92)(0.49,-0.11){1}{\line(1,0){0.49}}
\multiput(62.83,6.81)(0.49,-0.1){1}{\line(1,0){0.49}}
\multiput(63.31,6.71)(0.49,-0.09){1}{\line(1,0){0.49}}
\multiput(63.8,6.61)(0.49,-0.09){1}{\line(1,0){0.49}}
\multiput(64.29,6.52)(0.49,-0.08){1}{\line(1,0){0.49}}
\multiput(64.79,6.44)(0.49,-0.07){1}{\line(1,0){0.49}}
\multiput(65.28,6.37)(0.49,-0.07){1}{\line(1,0){0.49}}
\multiput(65.77,6.3)(0.5,-0.06){1}{\line(1,0){0.5}}
\multiput(66.27,6.25)(0.5,-0.05){1}{\line(1,0){0.5}}
\multiput(66.76,6.19)(0.5,-0.04){1}{\line(1,0){0.5}}
\multiput(67.26,6.15)(0.5,-0.04){1}{\line(1,0){0.5}}
\multiput(67.76,6.11)(0.5,-0.03){1}{\line(1,0){0.5}}
\multiput(68.26,6.09)(0.5,-0.02){1}{\line(1,0){0.5}}
\multiput(68.75,6.06)(0.5,-0.01){1}{\line(1,0){0.5}}
\multiput(69.25,6.05)(0.5,-0.01){1}{\line(1,0){0.5}}
\put(69.75,6.04){\line(1,0){0.5}}
\multiput(70.25,6.04)(0.5,0.01){1}{\line(1,0){0.5}}
\multiput(70.75,6.05)(0.5,0.01){1}{\line(1,0){0.5}}
\multiput(71.25,6.06)(0.5,0.02){1}{\line(1,0){0.5}}
\multiput(71.74,6.09)(0.5,0.03){1}{\line(1,0){0.5}}
\multiput(72.24,6.11)(0.5,0.04){1}{\line(1,0){0.5}}
\multiput(72.74,6.15)(0.5,0.04){1}{\line(1,0){0.5}}
\multiput(73.24,6.19)(0.5,0.05){1}{\line(1,0){0.5}}
\multiput(73.73,6.25)(0.5,0.06){1}{\line(1,0){0.5}}
\multiput(74.23,6.3)(0.49,0.07){1}{\line(1,0){0.49}}
\multiput(74.72,6.37)(0.49,0.07){1}{\line(1,0){0.49}}
\multiput(75.21,6.44)(0.49,0.08){1}{\line(1,0){0.49}}
\multiput(75.71,6.52)(0.49,0.09){1}{\line(1,0){0.49}}
\multiput(76.2,6.61)(0.49,0.09){1}{\line(1,0){0.49}}
\multiput(76.69,6.71)(0.49,0.1){1}{\line(1,0){0.49}}
\multiput(77.17,6.81)(0.49,0.11){1}{\line(1,0){0.49}}
\multiput(77.66,6.92)(0.48,0.12){1}{\line(1,0){0.48}}
\multiput(78.15,7.03)(0.48,0.12){1}{\line(1,0){0.48}}
\multiput(78.63,7.15)(0.48,0.13){1}{\line(1,0){0.48}}
\multiput(79.11,7.29)(0.48,0.14){1}{\line(1,0){0.48}}
\multiput(79.59,7.42)(0.48,0.14){1}{\line(1,0){0.48}}
\multiput(80.07,7.57)(0.48,0.15){1}{\line(1,0){0.48}}
\multiput(80.54,7.72)(0.47,0.16){1}{\line(1,0){0.47}}
\multiput(81.01,7.88)(0.47,0.17){1}{\line(1,0){0.47}}
\multiput(81.48,8.04)(0.47,0.17){1}{\line(1,0){0.47}}
\multiput(81.95,8.21)(0.47,0.18){1}{\line(1,0){0.47}}
\multiput(82.42,8.39)(0.23,0.09){2}{\line(1,0){0.23}}
\multiput(82.88,8.58)(0.23,0.1){2}{\line(1,0){0.23}}
\multiput(83.34,8.77)(0.23,0.1){2}{\line(1,0){0.23}}
\multiput(83.8,8.97)(0.23,0.1){2}{\line(1,0){0.23}}
\multiput(84.25,9.18)(0.23,0.11){2}{\line(1,0){0.23}}
\multiput(84.7,9.39)(0.22,0.11){2}{\line(1,0){0.22}}
\multiput(85.15,9.61)(0.22,0.11){2}{\line(1,0){0.22}}
\multiput(85.59,9.83)(0.22,0.12){2}{\line(1,0){0.22}}
\multiput(86.04,10.06)(0.22,0.12){2}{\line(1,0){0.22}}
\multiput(86.47,10.3)(0.22,0.12){2}{\line(1,0){0.22}}
\multiput(86.91,10.55)(0.22,0.13){2}{\line(1,0){0.22}}
\multiput(87.34,10.8)(0.21,0.13){2}{\line(1,0){0.21}}
\multiput(87.77,11.06)(0.21,0.13){2}{\line(1,0){0.21}}
\multiput(88.19,11.32)(0.21,0.14){2}{\line(1,0){0.21}}
\multiput(88.61,11.59)(0.21,0.14){2}{\line(1,0){0.21}}
\multiput(89.02,11.87)(0.21,0.14){2}{\line(1,0){0.21}}
\multiput(89.43,12.15)(0.2,0.14){2}{\line(1,0){0.2}}
\multiput(89.84,12.44)(0.2,0.15){2}{\line(1,0){0.2}}
\multiput(90.24,12.73)(0.13,0.1){3}{\line(1,0){0.13}}
\multiput(90.64,13.03)(0.13,0.1){3}{\line(1,0){0.13}}
\multiput(91.03,13.34)(0.13,0.1){3}{\line(1,0){0.13}}
\multiput(91.42,13.65)(0.13,0.11){3}{\line(1,0){0.13}}
\multiput(91.81,13.97)(0.13,0.11){3}{\line(1,0){0.13}}
\multiput(92.19,14.29)(0.13,0.11){3}{\line(1,0){0.13}}
\multiput(92.56,14.62)(0.12,0.11){3}{\line(1,0){0.12}}
\multiput(92.93,14.95)(0.12,0.11){3}{\line(1,0){0.12}}
\multiput(93.3,15.29)(0.12,0.11){3}{\line(1,0){0.12}}
\multiput(93.66,15.64)(0.12,0.12){3}{\line(1,0){0.12}}
\multiput(94.01,15.99)(0.12,0.12){3}{\line(0,1){0.12}}
\multiput(94.36,16.34)(0.11,0.12){3}{\line(0,1){0.12}}
\multiput(94.71,16.7)(0.11,0.12){3}{\line(0,1){0.12}}
\multiput(95.05,17.07)(0.11,0.12){3}{\line(0,1){0.12}}
\multiput(95.38,17.44)(0.11,0.13){3}{\line(0,1){0.13}}
\multiput(95.71,17.81)(0.11,0.13){3}{\line(0,1){0.13}}
\multiput(96.03,18.19)(0.11,0.13){3}{\line(0,1){0.13}}
\multiput(96.35,18.58)(0.1,0.13){3}{\line(0,1){0.13}}
\multiput(96.66,18.97)(0.1,0.13){3}{\line(0,1){0.13}}
\multiput(96.97,19.36)(0.1,0.13){3}{\line(0,1){0.13}}
\multiput(97.27,19.76)(0.15,0.2){2}{\line(0,1){0.2}}
\multiput(97.56,20.16)(0.14,0.2){2}{\line(0,1){0.2}}
\multiput(97.85,20.57)(0.14,0.21){2}{\line(0,1){0.21}}
\multiput(98.13,20.98)(0.14,0.21){2}{\line(0,1){0.21}}
\multiput(98.41,21.39)(0.14,0.21){2}{\line(0,1){0.21}}
\multiput(98.68,21.81)(0.13,0.21){2}{\line(0,1){0.21}}
\multiput(98.94,22.23)(0.13,0.21){2}{\line(0,1){0.21}}
\multiput(99.2,22.66)(0.13,0.22){2}{\line(0,1){0.22}}
\multiput(99.45,23.09)(0.12,0.22){2}{\line(0,1){0.22}}
\multiput(99.7,23.53)(0.12,0.22){2}{\line(0,1){0.22}}
\multiput(99.94,23.96)(0.12,0.22){2}{\line(0,1){0.22}}
\multiput(100.17,24.41)(0.11,0.22){2}{\line(0,1){0.22}}
\multiput(100.39,24.85)(0.11,0.22){2}{\line(0,1){0.22}}
\multiput(100.61,25.3)(0.11,0.23){2}{\line(0,1){0.23}}
\multiput(100.82,25.75)(0.1,0.23){2}{\line(0,1){0.23}}
\multiput(101.03,26.2)(0.1,0.23){2}{\line(0,1){0.23}}
\multiput(101.23,26.66)(0.1,0.23){2}{\line(0,1){0.23}}
\multiput(101.42,27.12)(0.09,0.23){2}{\line(0,1){0.23}}
\multiput(101.61,27.58)(0.18,0.47){1}{\line(0,1){0.47}}
\multiput(101.79,28.05)(0.17,0.47){1}{\line(0,1){0.47}}
\multiput(101.96,28.52)(0.17,0.47){1}{\line(0,1){0.47}}
\multiput(102.12,28.99)(0.16,0.47){1}{\line(0,1){0.47}}
\multiput(102.28,29.46)(0.15,0.48){1}{\line(0,1){0.48}}
\multiput(102.43,29.93)(0.14,0.48){1}{\line(0,1){0.48}}
\multiput(102.58,30.41)(0.14,0.48){1}{\line(0,1){0.48}}
\multiput(102.71,30.89)(0.13,0.48){1}{\line(0,1){0.48}}
\multiput(102.85,31.37)(0.12,0.48){1}{\line(0,1){0.48}}
\multiput(102.97,31.85)(0.12,0.48){1}{\line(0,1){0.48}}
\multiput(103.08,32.34)(0.11,0.49){1}{\line(0,1){0.49}}
\multiput(103.19,32.83)(0.1,0.49){1}{\line(0,1){0.49}}
\multiput(103.29,33.31)(0.09,0.49){1}{\line(0,1){0.49}}
\multiput(103.39,33.8)(0.09,0.49){1}{\line(0,1){0.49}}
\multiput(103.48,34.29)(0.08,0.49){1}{\line(0,1){0.49}}
\multiput(103.56,34.79)(0.07,0.49){1}{\line(0,1){0.49}}
\multiput(103.63,35.28)(0.07,0.49){1}{\line(0,1){0.49}}
\multiput(103.7,35.77)(0.06,0.5){1}{\line(0,1){0.5}}
\multiput(103.75,36.27)(0.05,0.5){1}{\line(0,1){0.5}}
\multiput(103.81,36.76)(0.04,0.5){1}{\line(0,1){0.5}}
\multiput(103.85,37.26)(0.04,0.5){1}{\line(0,1){0.5}}
\multiput(103.89,37.76)(0.03,0.5){1}{\line(0,1){0.5}}
\multiput(103.91,38.26)(0.02,0.5){1}{\line(0,1){0.5}}
\multiput(103.94,38.75)(0.01,0.5){1}{\line(0,1){0.5}}
\multiput(103.95,39.25)(0.01,0.5){1}{\line(0,1){0.5}}

\put(9,40){\makebox(0,0)[cc]{$r=0, z=0$}}

\linethickness{0.3mm}
\put(36,0){\line(0,1){80}}
\linethickness{0.3mm}
\put(36,40){\line(1,0){95}}
\put(30,80){\makebox(0,0)[cc]{$z$}}

\put(130,35){\makebox(0,0)[cc]{$r$}}

\put(70,45){\makebox(0,0)[cc]{$2L_0$}}

\end{picture}

}

In the previous sections we have discussed how by creating large  regions in space-time where
the brane tension can be made small we can ensure 
that a loop of the brane is not hidden by an event horizon.
In order to study the properties of an isolated flat brane we also need to be able to
make the size of the loop
arbitrarily large so that locally a portion of the brane is indistinguishable from that of an
infinite flat brane. This can be done using the same scaling transformation that was used in
\cite{2502.07883,2503.00601,2506.13876} to create an arbitrarily
large region in space-time where the moduli take 
values different
from their asymptotic values. However, since for codimension two branes
the metric and various scalar fields vary
logarithmically as a function of the radial distance from the brane, one might worry if there
is any violation of the scaling property due to this logarithmic 
dependence\cite{1011.5120}. In this section we
shall discuss why this is not the case, at least for the BPS branes that we have discussed
above.

For definiteness we shall illustrate this using the solution describing a fundamental string
in four dimensional heterotic or type II string theory. The solution describing the infinitely long 
string was given in \cite{Dabholkar} and may be
expressed as,
\be \label{estringsol}
ds_c^2 = -dt^2 + dz^2 + g_4^{-2}
\left(1 - {g_4^2\over 2\pi} \,  \ln {r\over L_0}\right) (dr^2+r^2d\theta^2), \qquad
\tau =i\, g_4^{-2}\left(1 - {g_4^2\over 2\pi} \,  \ln {r\over L_0} e^{i\theta}\right) \, ,
\ee
where $ds_c^2$ is the four dimensional Einstein frame metric and $\tau$ is the axion-dilaton field
whose real part measures the scalar obtained by dualizing the two form gauge field and
whose imaginary part measures the inverse of the square of the string coupling.
$g_4$ and $L_0$ are parameters of the solution. We shall take $g_4$ to be small but finite.
We shall see that the parameter $L_0$ is associated with the freedom of scaling transformation
described in \refb{escale}.
As expected, $\tau$ is shifted by 1 as we go around the
string and $\theta$ changes by $2\pi$. Near the core $r\to 0$, $\tau\to i\infty$ and hence the string
coupling is weak, but the solutions hits a singularity when ${g_4^2\over 2\pi} \,  \ln {r\over L_0}=1$.
For the string loop solution of the type we shall consider, we shall never hit this singularity.

\begin{figure}
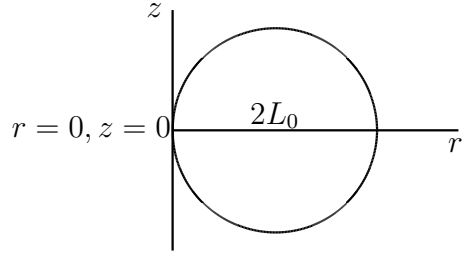


\begin{center}

\figstringloop

\end{center}

\caption{A string loop of coordinate radius $L_0$ in the $\theta=0$ plane.
\label{fstringloop}
}

\end{figure}

Now instead of considering an infinitely long string, let us consider 
a loop of  coordinate radius $L_0$. For definiteness, let us take the string to be in the $\theta=0$ plane,
passing through the point $r=0$, as shown in Fig.~\ref{fstringloop}.
Then for $r<<L_0$, $|z|<< L_0$, 
the solution will look approximately  as in \refb{estringsol}, 
but the logarithmic growth for
large $r$ gets
cut off at $r\sim L_0$. 
Beyond this distance the scalar fields
will evolve according to the three dimensional Laplace equation and approach a constant up to
corrections that fall off as inverse power of the distance from the loop. 
As a result 
the  metric at the center of the loop 
and at infinity will be approximately $\eta_{\mu\nu}$ after a rescaling of $r$ by $g_4$, 
and the imaginary part $\tau_2$ of 
$\tau$ at the center of the loop and at infinity will be approximately $g_4^{-2}$.
The solution will also evolve in time over a time scale of order $L_0$.

We now apply the scaling transformation on this solution
as follows.
In any general coordinate invariant theory in $D$ space-time dimensions,
where each term in the action has two derivatives, there is a
standard scaling property under which
a covariant tensor field $C^{(k)}$ of rank $k$, a contravariant tensor field 
$B^{(k)}$ of rank $k$ and
the action $S$ scales as follows\cite{9707207,2506.13876}:
\be\label{escale}
C^{(k)}\to \lambda^k C^{(k)}, \qquad B^{(k)}\to \lambda^{-k} B^{(k)}, \qquad
S\to \lambda^{D-2} S\, .
\ee
In particular the metric $g_{\mu\nu}$ 
scales as $\lambda^2$ and the scalars remain invariant.
This converts \refb{estringsol} to
\be \label{estringsolscaled}
ds_c^2 = -\lambda^2 dt^2 + \lambda^2 dz^2 + \lambda^2 g_4^{-2}
\left(1 - {g_4^2\over 2\pi} \,  \ln {r\over L_0}\right) (dr^2+r^2d\theta^2), \qquad
\tau =i\, g_4^{-2}\left(1 - {g_4^2\over 2\pi} \,  \ln {r\over L_0} e^{i\theta}\right) \, ,
\ee
Since the metric is scaled by an
overall constant, the new solution does not have any event horizon since the original
solution did not have one.
Under this scaling, the interval  over which the solution appears to be a straight 
string scales as
$\lambda$ and this can be taken to be as large as we like by taking
$\lambda$ large. Also the physical time over which the solution evolves will scale as $\lambda$,
showing that for large $\lambda$ the evolution slows down.
We can introduce new coordinates
$(t',z',r')=\lambda(t,z,r)$ so that the asymptotic
metric is given by $\eta_{\mu\nu}$ after a further rescaling of $r'$ by $g_4$. 
This effectively replaces $L_0$ by $\lambda\, L_0$.
In the new coordinate the string will appear to be straight in the region $|z'|<<\lambda L_0$
and $|r'|<<\lambda L_0$. 
Similarly the time interval $\Delta t'$ over which
the solution evolves will also be large, carrying a factor of $\lambda$.
Finally $\tau_2$  is still of order $g_4^{-2}$ at
infinity and 
at the center of the loop and $\tau_1$
changes by 1 as we go once around the string, showing
that the new solution still represents a single fundamental string.\footnote{By contrast, the
number of branes
of other codimensions scale by some power of $\lambda$ under the 
scaling\cite{2506.13876}.}
This allows us to study the properties of a straight string of any desired length provided
we have enough resources to produce a large string loop of this type.

One can make this discussion more explicit as follows. Let us consider a circular
loop of radius $L$ at rest at $x^0=0$ and let it evolve in time. If we take the string to lie in
the $x^1$-$x^2$ plane centered at $x^1=0$, $x^2=0$, 
then the subsequent evolution of the classical string is given in the
parametric form as\cite{vilenkin2}:
\be
X^0=L\, \xi, \quad X^1 = L\, \cos\xi\cos\sigma, \quad X^2 = L\, \cos\xi\sin\sigma\, .
\ee
In this coordinate system, the scaling transformation \refb{escale}  can also be described as
the scaling of $L$ by $\lambda$ keeping the metric fixed.
Now suppose that we want to study the properties of the string by staying at a distance $\ell_0$ from
the string for a period $t_0$. Our goal will be to show that for any given $\ell_0$ and $t_0$,
we can ensure that the displacement of the string remains small compared to $\ell_0$ during the
time $t_0$ so that the string appears static during the experiment.
For this we begin the experiment at $x^0=0$ at the point $x^2=0$, $x^1 = L-\ell_0$
so that the point on the string that is nearest to the experimentalist corresponds to $x^1=L$,
$x^2=0$. At $x^0=t_0$ this will reach the point
\be
\xi=t_0/L
\quad \Rightarrow\quad x^1=L\cos(t_0/L) \simeq L - t_0^2 / (2 L)\,  .
\ee
This shows that however large a time $t_0$ we need for the experiment, by taking $L$ to be
sufficiently large we can ensure that the string remains almost static during the experiment.
In particular, to measure the monodromy around the string remaining at a
distance $\ell_0$ we need a time of order $2\pi \ell_0$, during which the string can be made
to remain almost static, provided $\ell_0$ does not scale with $L$.

Even though the argument based on the scaling transformation is correct, it is instructive to examine
in some detail possible sources of logarithmic violation since this has potential consequences 
for conjectured non-BPS branes for which the analog of the solution 
\refb{estringsol} is not known. For this we write the expression for $\tau$ given
in the scaled coordinates described below \refb{estringsolscaled} as
\be
\tau = i g_4^{-2} - {i\over 2\pi} \ln {w\over L}, \qquad w\equiv r'\, e^{i\theta}, \quad L\equiv 
\lambda \, L_0 \, ,
\ee
and try to compute the energy per unit length of the string
due to the scalar kinetic term by taking an upper cut-off of $L$
on $|w|$. 
If $\tau$ had been a canonically normalized scalar, then the energy of the field per unit
length
will be proportional to
\be \label{enaiveenergy}
\int d^2 w \, \p_w\tau \p_{\bar w} \bar\tau = {1\over 2\pi} \int dr' \, r^{\prime -1} \, .
\ee
This requires both an upper and a lower cut-off to get a finite answer.
The upper cut-off is the size $L$ of the loop since beyond this distance the scalar field
will evolve according to the three dimensional Laplace equation and approach a constant up to
corrections that fall off as inverse power of the distance from the loop. 
The lower cut-off may be taken to be
some microscopic scale $\mu$. This gives the energy per unit length
contained in the scalar field to be
$\ln (L/\mu)$ and the total energy to be $L\ln (L/\mu)$, contradicting the claim that the 
energy scales as $L$\cite{1011.5120}. The scaling argument fails since the microscopic
cut-off implicitly requires terms in the action other than those containing only two derivatives.

In actual practice, the kinetic term of $\tau$ has an additional factor of $\tau_2^{-2}$.
With this the energy per unit length contained in the scalar field becomes proportional to
\be\label{ecorrectenergy}
\int d^2 w \, \tau_2^{-2} \, \p_w \tau \p_{\bar w} \bar\tau = {1\over 2\pi} 
\int dr' \, r^{\prime -1} \left( g_4^{-2} - {1\over 2\pi} \ln {r'\over L}\right)^{-2} 
= \left( g_4^{-2} - {1\over 2\pi} \ln {r'\over L}\right)^{-1} \Bigg|_{\mu}^L \, .
\ee
Up to a numerical factor,
$g_4$ can be equated to the asymptotic
value of the string coupling. Then \refb{ecorrectenergy} 
may be written as
\be
g_4^2 - \left(g_4^{-2} + {1\over 2\pi} \ln{L\over \mu}\right)^{-1}\, .
\ee
We now see that the second term vanishes in the limit 
$\mu\to 0$ leaving us with the first term $g_4^2$. Therefore the result is independent of $L$, 
in agreement with the scaling laws.
The important observation here is that the result is not sensitive to the microscopic scale $\mu$.
Had it been otherwise, we could not have applied the scaling argument since the microscopic
cut-off scale will violate scaling laws.

The analysis described here can be generalized to other BPS
exotic branes of the type discussed in section \ref{s2}, \ref{s3} and \ref{s4}. 
In particular the scalar field
produced by these branes take the same form as \refb{estringsol}, 
with $\tau$ having different interpretation
for different branes. For example for the D7-brane $\tau$ is the axion-dilaton modulus of the ten
dimensional type IIB string theory, while for the NS 5-brane with two transverse directions
compactified on a $T^2$,
$\tau$ is the complexified Kahler 
modulus associated with the $T^2$. 
But in
all such cases the analogs of \refb{estringsol} and 
\refb{ecorrectenergy} hold, with $g_4$ having different interpretations, 
and ensures that the dependence on the
microscopic cut-off disappears, restoring the scaling law. This is easiest to see in type II string
theory on $T^6$ where the fundamental string is related to various exotic strings by various duality
transformations and we can construct the solution describing these exotic strings by duality
transformation of the solution \refb{estringsol}. Once these solutions have been constructed, a
subset of them carrying NSNS charges
can also be regarded as solutions in heterotic string theory on $T^6$.

To contrast this with general codimension two branes, let us compute the energy of a 
vortex string solution in a field theory of a complex
scalar field with potential $\lambda (\phi^*\phi-a^2)^2$.  In this case we can construct a 
solution with scalar field profile $\phi=f(r) e^{i\theta}$ with $f(r)\to a$ for $r>>\mu\equiv
1/ (a\sqrt \lambda)$ and $f(0)=0$. The scalar kinetic term then produces an 
energy per unit length
of the string of order
\be
\int 2\pi r dr \, (\p_r \phi^* \p_r \phi + r^{-2} \p_\theta\phi^*\p_\theta\phi)
\simeq 2\pi \int_\mu^L dr\, r^{-1} a^2 \simeq 2\pi |a|^2 \ln(L/\mu)\, ,
\ee
where we have put an upper cut-off $L$ on the $r$ integral by taking the string to be a
macroscopic loop of length $L$. We see that this violates the scaling law. This can be traced
to the dependence of the result on the microscopic cut-off scale $\mu$.
We have discussed this example here, since while studying the energy of macroscopic loops
of non-BPS branes, we need to ensure that such logarithmic dependence on the size $L$
of the loop is absent.

\sectiono{U-duality as gauge symmetries} \label{sudual}

It is generally believed that once we take into account the
effects of quantum mechanics and gravity, there are
no global symmetries. In this section we shall discuss the role of codimension two branes
in establishing that the U-duality symmetries of string theory are 
(spontaneously broken) discrete gauge symmetries. 
As in the rest of the paper,
our analysis will apply
only to asymptotically flat space-times. 

Before discussing whether the duality transformations are global or gauge symmetries, let us
first discuss whether they are symmetries at all. At a generic point in the moduli space the
U-duality symmetries are spontaneously broken by the asymptotic values of the 
moduli. Nevertheless, as discussed in \cite{2502.07883,2503.00601,2506.13876}, 
in the class of theories we are analyzing,  given
any set of asymptotic values of the moduli collectively labelled as A, we can create an
arbitrarily large region $\RR$ of space-time in which the moduli take any other set of values B.
Therefore, if we choose B and A to be related by a U-duality transformation then U-duality symmetry
predicts the result of any
experiment in the region $\RR$  by knowing the corresponding results in the
asymptotic region, This is clearly a testable prediction and identifies U-duality
transformations as spontaneously broken symmetries.

In order to distinguish between gauge and
global symmetries, we shall take a practical approach as seen from an experimentalist's viewpoint
instead of the more formal and rigorous approach adopted in \cite{1810.05337,1810.05338}.
Given a  transformation $g$, we shall say that it is a gauge symmetry if two field
configurations related by $g$ are identified. Hence we should allow classical configurations
(coherent states) with the property that in that state,
as we go around a closed loop, the fields undergo a $g$ 
transformation\cite{0510033,1011.5120}.\footnote{This
is sufficient to establish that a given transformation is a gauge symmetry, but there may be
other ways to do this. 
Nevertheless it is a very useful tool, since in a theory of gravity measurements are
always more subtle due to the possibility of black hole formation. For example a measurement
that requires us to simultaneously place finite mass
detectors covering the whole solid angle around a point
may not be possible since such a system of detectors will have a large enough mass so that
they form a blackhole whose horizon surrounds them.}
Conversely, the
existence of such a state in the theory will imply that the transformation under consideration is a
gauge symmetry. In this context we also note that whether a given transformation is a gauge symmetry
or not could depend on the string compactification we have, -- establishing that a transformation is a
gauge symmetry in one compactification does not establish that it is a gauge symmetry in another
compactification unless in the latter theory 
we can create an arbitrarily large region in space-time inside which we have the former theory.

We shall now try to make this requirement more precise.
A  transformation $g$ can be called a gauge symmetry if 
the system has a (time dependent)  state that has two time-like
curves $C_1$ and $C_2$, beginning and ending at the same points $A$ and $B$ 
in space-time, with the following properties:
\begin{enumerate}
\item  Both curves lie in regions of space-time where,
within the desired precision, the environment is locally
indistinguishable 
from the vacuum associated with the values of the moduli
scalar fields in that region.
In particular the components 
of the stress tensor in the rest frame of the particle moving along $C_1$ or $C_2$, 
multiplied by the
proper time along the trajectory, must be small so that the integrated
effect of departure from the vacuum along the trajectories remain small. 
This does not exclude (scalar) fields that vary along the trajectory and
undergo monodromy, but
the stress tensor produced by the varying fields must satisfy the requirement
described above.

\item Since the trajectories may have to undergo acceleration, 
we need to ensure that the  
acceleration measured in the rest frame of the particle is small so 
that a test particle moving along
such a trajectory effectively behaves as if it is moving in vacuum. 
This will also ensure that the effect of Unruh radiation experienced by a particle
moving along the trajectory remains small.
\item If we take a localized quantum state at $A$, e.g. a particle, and 
transport it to $B$, then the state at $B$ transported along $C_1$ differs from the state at
$B$ transported along $C_2$  by the transformation $g$.
\end{enumerate}
Conditions 1 and 2
on the paths are important since they ensure that the transformation by $g$ is not due to
any other dynamical effect like  scattering with another particle.

First we review the case of a quantum field
theory without gravity. 
Let us take the rigid part of the $U(1)$ gauge symmetry in 3+1 dimensions
under which a 
charged particle picks up a phase. In this case we can consider a solenoid carrying a
magnetic field and $C_1$ and $C_2$ can be taken to be time-like curves from $A$ to $B$
on two sides of the magnetic flux lines.
If we transport a quantum state at $A$ to $B$, then the state transported along $C_2$ and a 
state transported along $C_1$ will have a relative phase
due to Aharanov-Bohm 
effect even though neither path encounters any magnetic field and locally experiences vacuum.
Of course when full quantum effects are taken into account the effect of the
flux will be felt outside the solenoid and neither path will be really in the vacuum, but 
by taking the size of the system to be very large keeping the cross section of the solenoid
fixed and the paths to be far away from the 
solenoid, we can reduce this effect to any desired degree.

Once the effect of gravity is taken into account there are further complications. First of all,
due
to the long range nature of gravity and the constraints,
even classically we cannot switch off the gravitational field
outside a given region carrying non-vanishing energy density.
Second, since the metric fluctuates, there is no gauge 
invariant notion of a fixed
curve in space-time. 
For this reason the conditions 1 and 2 above become more important, since
only if the local environment of every point on the trajectory  is approximately 
that of the vacuum, we can ignore the classical background gravitational field and the effects
of quantum
gravitational fluctuations for low energy processes.
This can be
achieved 
by using the scaling transformation by $\lambda$ described in \refb{escale} to 
generate a new solution from a
given solution and maps classical particle trajectories to new trajectories. 
Under this scaling the proper time along the trajectory scales as
$\lambda$ while 
all the scalars constructed from $2n$ derivatives of fields scale as $\lambda^{-2n}$ and
becomes small for large $\lambda$ for $n\ge 1$. With a little work one can also show that
the energy momentum tensor $T_{\mu\nu}$ (including the gravitational contribution), 
measured in the rest frame of the particle in which the local metric is $\eta_{\mu\nu}$, the
first derivative of the metric is zero and
the particle is at rest, scales as $\lambda^{-2}$ in the large $\lambda$ limit.
Therefore not only locally the background appears to be that of the vacuum, but the integrated
effect of the departure from the vacuum over the entire trajectory falls at least as fast as
$\lambda^{-1}$ for large $\lambda$ 
and hence vanishes in the $\lambda\to\infty$ limit. One can also show that the acceleration
of the particle at any point on the trajectory, measured in the rest frame of the particle, scales
as $\lambda^{-1}$. Hence the Unruh temperature scales as $\lambda^{-1}$ and the Unruh radiation
density scales as $\lambda^{-D}$. So even 
the integrated effect of Unruh radiation encountered by the particle
vanishes as $\lambda^{1-D}$, and the particle moving along the trajectory can be made
to feel that it is in the vacuum to any desired degree of accuracy. 
Since the particle always sees the local environment as the
vacuum, we expect that the semiclassical notions like particle trajectories still make sense.

One can now see how the existence of $(s,r)$ strings in heterotic string theory in
$T^6\times \RRR^{3,1}$ and $(s,r)$ seven branes in type IIB string theory in
$\RRR^{9,1}$ can be used to show
that the corresponding SL(2,$\ZZZ)$ symmetries in these theories are gauge symmetries.
By taking $C_1-C_2$ to be a curve that links the world-volume of the string or the
seven brane, one sees that the state transported along $C_1$ and along $C_2$ differ by the
SL(2,$\ZZZ)$ transformation \refb{emonodromy}. Furthermore by using the scaling transformation
\refb{escale} and taking $\lambda$ to be large one can ensure that the deviation from
vacuum along the trajectory remains small. This establishes that SL(2,$\ZZZ)$ transformations
of the form \refb{emonodromy} are gauge symmetries.
This is sufficient for establishing that all SL(2,$\ZZZ$) transformations are gauge
symmetries, since 
the matrices $\pmatrix{1 & 1\cr 0 & 1}$ and $\pmatrix{1 & 0 \cr -1 & 1}$, -- the monodromy
matrices associated with the (1,0) and (0,1) strings, --  generate the
whole S-duality group. This does not mean that we can construct a single string
or seven brane  that produces
the desired SL(2,$\ZZZ)$ monodromy. However by threading a path through different $(s,r)$ 
string / seven brane
loops, situated in different regions in space-time where their tensions are small,
we can produce the desired monodromy.

A similar analysis can be done for the T-duality symmetries in 
heterotic string theory on $T^6\times \RRR^{3,1}$.
For example an NS 5-brane wrapped on four of the six
directions of $T^6$ will produce a monodromy that shifts the component of the 2-form field
along the other two directions of the torus by one unit. Various other (exotic) 
codimension two branes
related to this by T-duality trasformation produces other monodromies in the T-duality group,
and together they establish that the whole T-duality group is a 
spontaneously broken discrete gauge symmetry. 
More generally in any string compactification,
monodromy around the exotic branes can be used to establish that the discrete U-duality
groups of those theories are also (spontaneously broken) gauge symmetries.

Note that in the heterotic string theory on $T^6\times \RRR^{3,1}$,
as we transport a probe charged particle around an $(s,r)$ string, the charge
carried by the probe changes by the monodromy transformation. Since the total charge
is conserved, the extra charge must be deposited on the string. A simple example of this
is the transport of a KK monopole around a fundamental 
string. By the rules of duality  transformation, the KK monopole picks up one
unit of fundamental string winding charge along the compact direction. This is achieved
by the macroscopic fundamental string undergoing a wrapping along the compact
direction during the transport of the KK monopole around it. A general $(s,r)$ string
will pick both electric and magnetic charges under such processes. 
Their ability to absorb the charge is encoded in the existence of
appropriate zero modes on these strings, whose excitations allow the string to carry the
charges that they are expected to absorb during the transport of dual charged particles around
them\cite{9302038}.

\sectiono{AdS vs flat space-time} \label{sads}

The results of the previous section may appear to contradict what we know from
anti-de Sitter space-time where the spontaneously broken gauge symmetries in the
bulk theory of gravity have no natural manifestation in the dual boundary 
theory\cite{1810.05337,1810.05338}. This can be
illustrated in the celebrated example of the duality between N=4 supersymmetric Yang-Mills
theory and type IIB string theory on $AdS_5\times S^5$\cite{9711200}.
The type IIB string theory has
SL(2,$\ZZZ)$ S-duality symmetry that is spontaneously broken at a generic point in the moduli
space. In the dual boundary theory this is reflected in the fact that the $\NN=4$
supersymmetric Yang-Mills theories at two different values of the complex 
coupling $\tau$ are
equivalent if the corresponding values of $\tau$ are related by SL(2,$\ZZZ)$ 
transformation\cite{9812012,2208.09396}. 
Hence for a generic value of $\tau$ the duality symmetry is explicitly broken
in the boundary theory, and does not lead to any direct experimental consequences.
So if an experimentalist in the bulk can verify the existence of this symmetry by
measuring monodromy around an $(s,r)$ 7-brane, this would lead to a contradiction.

The resolution of this puzzle seems to lie in the difference in our ability to measure
the monodromy around a 7-brane in flat space-time and in AdS space-time. In flat
space-time, a
7-brane loop of size $L$ can last for a time of order $L$ that can be made as large as
we want.  In AdS space-time the scaling `symmetry' \refb{escale} is broken due to the
presence of the cosmological constant, and cannot be used to generate new solutions from
old ones. More specifically, in AdS space-time such a seven brane loop  will collapse
in a time scale set by the AdS scale\cite{2506.13876}. 
Hence we cannot do a controlled experiment 
on the seven branes for arbitrarily large time and determine the monodromy around them
to the desired level of confidence. In fact,  in AdS space-time
we do not even have a controlled test of the duality transformations being symmetries at all, since
we cannot create arbitrarily large spatial regions, lasting for arbitrarily long time, inside which the
moduli take values that are different from their asymptotic values\cite{2506.13876}.
This leads to a consistent conclusion 
that the duality symmetry does not
have any observable consequences either in the boundary theory or in the bulk theory.
Nevertheless, for $SU(N)$ super-Yang-Mills theory with coupling $g_{YM}$, 
this property must emerge in the limit of large $N$ at fixed $g_{YM}$, since in this
limit $AdS_5\times S^5$ approaches flat space-time.

\sectiono{Non-BPS branes} \label{s6}

So far we have discussed BPS exotic branes. String theory also has conjectured
non-BPS exotic branes for which we do not know the tension. 
However if one accepts the criterion for the existence of a
codimension two brane 
described above, one can propose the following as the requirement  for the existence
of a non-BPS exotic brane:
\\
\noindent {\it A necessary criterion for a non-BPS exotic brane to exist in a theory is that its
tension is small enough in some corner of the moduli space so that a
large  spherical brane is not hidden
behind its Schwarzschild radius.}

We shall now discuss some examples of conjectured codimension two non-BPS branes.
The first example is the reflection seven brane (R7-brane) discussed in 
\cite{2212.05077,2302.00007,2507.21210,2509.03573}. These are
conjectured codimension two branes in type IIA and type IIB string theories with the 
property that
as we transport a state around such a brane, the state returns transformed by a $(-1)^{F_L}$
transformation. Here $(-1)^{F_L}$ is the $\ZZZ_2$ transformation that changes the signs of all the
$RNS$ and $RR$ states of the theory. The existence of these branes would establish in particular
that $(-1)^{F_L}$ is a gauge symmetry. From our earlier analysis of codimension two branes
it follows that in order for an R7-brane to exist in the strong sense described above, there must be
regions in the moduli space where the brane tension is sufficiently small 
so that a macroscopic loop
of the R7-brane will not be behind its own event horizon.\footnote{This is not just a condition on
the tension but also involves information on what other sources of massless fields the brane
provides, since whether the macroscopic loop will be behind the horizon is determined by a
combination of all these data.}
While this constraint is not as severe as the
one for codimension one brane, this still puts a strong restriction on the tension of the brane.

Since the R7-brane has a 
monodromy of $(-1)^{F_L}$, all RR fields must change sign as we go once around the loop.
This applies to the RR scalar field as well. So if we place an R7-brane loop in a background
where the RR scalar field takes a non-zero value $a_0$, 
then it must vary as we go around the R7-brane,
having the form $a_0 e^{i\theta/2}$ far away from the brane. 
If the dilaton remains constant, then the kinetic term of
the RR scalar will give an energy of order \refb{enaiveenergy}, leading to scaling violation.
So either the dilaton must vary, tempering the divergence as in \refb{ecorrectenergy}, or we
have to conclude that a macroscopic loop of R7-brane can only exist in regions of
space-time where the RR scalar vanishes.

As a side remark, we note that the existence of R7-brane by itself is not needed for establishing
that $(-1)^{F_L}$ is a gauge symmetry. What one needs is a combination of 7-branes, possibly
existing in different regions of space-time with different values of moduli, such that we can find a
path looping around these branes along which the total monodromy is $(-1)^{F_L}$. For example
the monodromy around a D7-brane, acting on the NSNS 2-form field and RR 2-form field,
is given by $\pmatrix{1 & 1\cr 0 & 1}$. So if we find another brane that produces a monodromy
$\pmatrix{1 & -1\cr 0 & -1}$ then the product of the two will give
\be
\pmatrix{1 & -1\cr 0 & -1} \pmatrix{1 & 1\cr 0 & 1}= \pmatrix{1 & 0\cr 0 & -1}\, ,
\ee
which is precisely $(-1)^{F_L}$ transformation. Therefore the 
existence of a brane with monodromy
$\pmatrix{1 & -1\cr 0 & -1}$ will also be sufficient to establish that $(-1)^{F_L}$ is a gauge
symmetry.

Another example of a $\ZZZ_2$ symmetry in string theory, that is not obviously a gauge
symmetry,  is the exchange of two $E_8$
factors in the $E_8\times E_8$ heterotic string theory in ten space-time dimensions. 
If we had a (non-BPS) 7-brane with the
property that as we go around the 7-brane the two $E_8$ factors get exchanged and the
tension of the brane is sufficiently small in some corner of the moduli space so that a loop
of the brane is not hidden behind its event horizon, then it would establish that this symmetry
is a gauge symmetry. The world-sheet description of such branes was proposed in 
\cite{2303.17623,2411.04344}, but  it is not known if large loops of these branes
can avoid being hidden behind an event horizon.
Similar questions can
be asked for other known symmetries in string theory, {\it e.g.} the ones analyzed in
\cite{2507.12467}.

There is also a conjectured 
codimension one brane in ten dimensional string theory that is supposed to
separate the vacua of
type IIA and type IIB string theory\cite{1909.10355}. If we use the strong criteria of their
existence, namely that we should be able to create arbitrarily large size loops of these branes
so that it looks locally flat and study their properties, then it follows from the discussion in section
\ref{s1} that this brane exists only if its tension, measured in Planck units, vanishes in some corner
of the moduli space. Since this cannot happen in the weak coupling limit of either string theory
where we know the spectrum of the theory, this should happen at some finite value $g_0$ of the
coupling. In that case we can produce a large region in space-time where the coupling is
arbitrarily close to $g_0$ and study these branes. Vanishing of the tension at some point
in the moduli space is clearly a strong constraint on the brane.

\subsection{Infinite tension branes}

Refs. \cite{2303.17623,2411.04344}  proposed the existence of 
some infinite tension branes in ten
dimensional heterotic string theories. These have codimension $(n+1)$ with $n=4$ and $8$, i.e.
they are 4-branes and 0-branes. They are sources of 1-form gauge field, with the property 
that at a distance $r$ away from
the brane, the field strength falls off as $1/r^2$. Therefore the energy density falls off as
$1/r^4$ and the energy per unit volume of the brane has a term proportional to
\be 
\int^\infty  r^n d r \times r^{-4}\, .
\ee
This diverges for $n=4$ and $n=8$. 

Now, even for finite tension, infinite branes have infinite energy (except for 0-branes) and hence 
are not regular states in the theory.  Having infinite tension makes things worse.
However the  relevant question, both for finite and infinite tension branes, is
whether we can construct a spherical
brane of radius $L$  that has energy that grows slower than
$L^7$ for large $L$.\footnote{For 0-branes this constitutes
a brane anti-brane pair separated by a distance $2L$.}
In that case its Schwarzschild radius would grow slower than $L$ and the brane would
not be shielded by an event horizon.
We have already seen that for finite tension branes this condition is satisfied for branes
of codimension larger than two. For 
a spherical  infinite tension brane
of the type described in \cite{2303.17623,2411.04344}, the
gauge field strength will fall off as $r^{-2}$ up to a distance $L$ and fall off much faster
beyond this distance since the gauge field configuration on the celestial sphere becomes
topologically trivial.
Such  brane configuration  will
have total energy density proportional to
\be
\int^L r^8 \, d r \times  r^{-4} \sim L^5\, .
\ee
This shows that indeed the total energy grows slower than $L^7$ and  the 
would be Schwarzschild radius grows as $L^{5/7}$. Since for large $L$ this is less than $L$,
we see that the brane is not shielded from the asymptotic observer by an
event horizon. 

By taking $L$ sufficiently large we can ensure that
locally the brane looks like a flat brane to any desired accuracy.
Whatever property of a flat brane of this type we want
to study can now be studied by performing experiments  on such branes.

\sectiono{Exposing the exotic branes from compact dimensions} \label{s5}

So far we have discussed the possibility of studying macroscopic loops of exotic branes. However,
exotic branes are also present in many string
compactifications\cite{9602022,9602114,9603161,9611007,0406102}. 
These  fill the non-compact
part of the space-time but may be localized along some of the compact directions. 
At a generic point in the moduli space
of the theory the sizes of the transverse directions are small and we cannot isolate these branes
and study them. However in special regions of the moduli space where the transverse directions
become large, the exotic branes may get isolated from the rest of the system 
and one can perform experiments to study their
properties.

We note, however, that the existence of these branes 
cannot be used to argue that the monodromy around such a brane  is a gauge symmetry 
in the higher
dimensional theory that gets exposed. 
For this we have to be able to create loops of these
branes in the higher dimensional theory itself. From this perspective, the goal of this section
is somewhat different from that in the earlier sections. Instead of using exotic branes to
show that certain transformations are gauge symmetries, we simply explore exotic branes in
asymptotically weakly coupled string compactifications.

\subsection{(s,r) seven branes from F-theory compactification} \label{sftheory}

In this subsection we shall illustrate the procedure
in the context of one particular example: F-theory on
$K3\times T^4$\cite{9602022}.
We can describe this compactification
as type IIB string theory compactified on the product of a two dimensional sphere 
$B$ 
times a four dimensional torus $T^4$ of  finite size (in string scale), with twenty four
$(s,r)$ seven branes localized
at various points on the sphere $B$ and extending along 
$T^4\times \RRR^{3,1}$. 
When the size of the sphere $B$ becomes large then the 7-branes get exposed and
can be studied by an asymptotic observer. Our goal will be to examine how this can be
achieved in a theory where the size of $B$ is small asymptotically.

The theory described above
 is in the same moduli space as heterotic string theory on $T^6$. We shall first identify the
region in the moduli space in which the F-theory description is valid, 
For this we note that this theory also has a description as
a orientifold of type IIB on $T^2\times T^4$\cite{9605150}, 
where the orientifold group acts by reversing the sign
of the two directions along $T^2$ together with the internal symmetry $(-1)^{F_L}\Omega$ where
$(-1)^{F_L}$ is the symmetry that changes the sign of all the  RNS and RR states 
and $\Omega$ is the world-sheet
parity transformation. The sphere $B$ is identified as the space $T^2/\ZZZ_2$, where $\ZZZ_2$
acts by reversing the sign of the two directions of $T^2$.
This produces four orientifold seven planes
localized at the four fixed points on $T^2$. To cancel the RR charges carried by the orientifold
seven planes we need to add 16 D7-branes placed at various points on $T^2/\ZZZ_2$ and 
extending along $T^4\times \RRR^{3,1}$. Once non-perturbative effects are taken into account, each
orientifold plane splits into a pair of $(s,r)$ seven branes for appropriare $(s,r)$. This gives altogether
24 seven branes which can be identified as those required for an F-theory compactification.
If we place 4 D7-branes on each of the four orientifold planes, then the split in the orientifold
also disappears and we get $SO(8)^4$ unbroken gauge group, one from each orientifold plane.
We shall assume that the configuration is such that the $SO(8)^4$ is slightly broken to 
$U(1)^{16}$, so that the D7-branes are not exactly on top of the orientifold plane but they are
close enough so that the analysis of \cite{9605150} holds. In the limit where the size of 
$B=T^2/\ZZZ_2$
becomes large keeping the Wilson lines fixed (so that the ratios of gauge boson masses
remain fixed) the distances between the seven branes increase and we can study them
individually.

We shall now examine what this limit correspond to in the heterotic variables. In the type IIB
description, let $\wt R_{B}$ denote 
the order of the 
radii of the four circles of $T^4$ which we shall label by coordinates $x^4,x^5,x^6,x^7$
and let $R_{B}$ denote the order of the 
radii of the two circles of $T^2$ which we label by $x^8,x^9$, all measured
in the type IIB string metric:
\be
R_4{}_{B}\sim R_5{}_{B}\sim R_6{}_{B}\sim R_7{}_{B}\sim \wt R{}_{B}, \qquad R_8{}_{B}\sim R_9{}_{B}\sim R{}_{B}\, .
\ee 
Also let $g_{B}$
be the coupling constant of ten dimensional type IIB theory.\footnote{Note that in the F-theory limit
the type IIB coupling varies over the sphere $B$ except in the special case when each orientifold
plane has 4 D7-branes on it. When we consider the case where each orientifold plane has four
D7-branes near it, $g_B$ can be taken to be the value of the coupling away from the orientifold
D7-brane system.}
We shall follow 
the convention 
that a string coupling without superscript will always denote the ten dimensional string 
coupling while the four dimensional coupling will carry a superscript (4).
We now make T-duality 
transformation along the 8 and
9 directions to convert this into type I string theory on $T^2\times T^4$, with the gauge group 
SO(32)
broken to $U(1)^{16}$ by Wilson lines along the 8 and 9 directions.
The ten dimensional coupling constant $g_I$ of the type I string theory and 
the radii $R_I$ of the
8,9 directions and $\wt R_I$ of the 4,5,6,7 directions measured in the type I metric are given by
\be
\wt R_I\sim \wt R{}_{B}, \qquad R_I \sim  1/R{}_{B}, \qquad g_I \sim  g_{B} 
\, R{}_{B}{}^{-2}\, .
\ee
In ten dimensional Planck units, the radii $\wt r$ of the 4,5,6,7 directions and
$r$ of 8,9 directions are given by
\be
\wt r \sim  \wt R_I g_I{}^{-1/4}\sim \wt R{}_{B}  R{}_{B}{}^{1/2} 
g_{B}{}^{-1/4}\, , \qquad r \sim  R_I g_I{}^{-1/4} 
\sim  R{}_{B}{}^{-1/2} g_{B}{}^{-1/4}\, .
\ee
In the heterotic description the radii $\wt R_H$ of the 4,5,6,7 directions,
$R_H$ of 8,9 directions and the ten dimensional coupling $g_H$ are given by
\be \label{eftheory1}
g_H 
\sim  1/ g_I \sim  g_{B}{}^{-1}R{}_{B}{}^{2}, \quad
\wt R_H \sim  \wt r \, g_H{}^{1/4} \sim  \wt R{}_{B}  R{}_{B}{} g_{B}^{-1/2}\, , \quad
R_H \sim  r \, g_H{}^{1/4} \sim  g_{B}^{-1/2}\, .
\ee
From this we get the four dimensional heterotic string coupling to be
\be\label{eftheory2}
g^{(4)}_H\sim  g_H \wt R_H^{-2} R_H^{-1} \sim  g_{B}^{1/2} \wt R{}_{B}{}^{-2}\, .
\ee
Finally, inverting \refb{eftheory1}, \refb{eftheory2} we get
\be \label{eftheory3}
g_B\sim R_H^{-2}, \qquad \wt R_B\sim \left(g^{(4)}_H\right)^{-1/2} \, R_H^{-1/2}, 
\qquad R_B \sim \wt R_H \, R_H^{-1/2}\, \left(g^{(4)}_H\right)^{1/2} \, ,
\ee
As in the original description, 
the $SO(32)$ gauge group is broken to its $U(1)^{16}$ subgroup by Wilson lines along
the 8 and 9 directions.  

Since to expose the 7-branes in the F-theory description we need $R{}_{B}$ and $\wt R_B$ 
to be large, 
with $g{}_{B}$ finite, we see from \refb{eftheory1}, \refb{eftheory2} that this translates to
finite $R_H$, small $g^{(4)}_H$ and large $\wt R_H$. 
We shall now demonstrate how we can create a large region 
near the horizon of a
suitable black hole
where the moduli take such values,
in a theory where the
asymptotic moduli take finite values.\footnote{It was shown in 
\cite{2506.13876} that by considering a system
of nested black holes we can access any point in the moduli space of heterotic string theory on $T^6$
from any given point. Here the goal is to show that the F-theory configuration can be achieved by
a single black hole. In some sense, higher dimensional theories have less moduli and hence are
easier to reach.}
For simplicity we shall assume that the asymptotic moduli are such that the appropriate Wilson
lines along the 8 and 9 directions are already turned on and 4,5,6,7 directions are described by the
product of four circles at their self-dual radii, and are orthogonal to the 8 and 9 directions.
These assumptions are not necessary but will simplify the solution. We now consider a 
purely electrically charged black hole in this theory carrying momentum charges along the
4,5,6,7 directions. This corresponds to setting  in the solution
(A.1) of \cite{2503.00601}, $\beta=-\alpha$ and $\vec p=\vec n$ for the first
four components, with the rest of the components set to zero.  The quantity of interest is the
heterotic string metric $G$ on $T^6$, and the four dimensional string coupling $g^{(4)}_H$,
which are given by
\be \label{e9.6}
G = I_6 + { 2m\over \rho} \, \sinh^2\alpha\, nn^T, \qquad g^{(4)}_H= g^{(4)}_0\, \left(1+ 
{ 2m\over \rho} \, \sinh^2\alpha\right)^{-1/4}\, ,
\ee
near the horizon of the black hole. Here $\rho$ is the radial coordinate, $g^{(4)}_0$ is the
asymptotic value of the heterotic string coupling, and $m$, $\alpha$
and the six dimensional unit vector $\vec n$ are parameters labelling the mass and charges
of the solution. The horizon is situated at $\rho=2m$, and we are considering a region where
$\rho\sim m$ but $\rho>2m$ so that the region is outside the horizon.
Following \cite{2503.00601} we take $\alpha$ large, corresponding to a near extremal black hole,
and 
\be
\vec n = (1, N^{-1}, N^{-2}, N^{-3}, 0, 0) / \sqrt{1+N^{-2}+N^{-4}+N^{-6}}\, ,
\ee
for some large integer $N$.
Consider now a cycle in $T^6$ that joins the point $(0,0,0,0,0,0)$ to $2\pi 
(k_1,k_2,k_3,k_4,k_5,k_6)$
in the $(x^4,x^5,x^6,x^7,x^8, x^9)$ directions for some integers $k_1,k_2,k_3,k_4, k_5,k_6$. 
The length of this cycle is given by,
\be
L(\vec k)=2\pi \sqrt{\vec k^2 +  { 2m\over \rho} \, \sinh^2\alpha\, ( \vec k.\vec n)^2}\, .
\ee
Let us choose $\rho$, $\alpha$ and $N$ such that
\be \label{e5.12}
{ 2m\over \rho} \, \sinh^2\alpha\, = N^{8} \, .
\ee
This can be achieved by taking $\alpha$ to be large, which translates
to the black hole being near extremal.
Then we get 
\be
L(\vec k) = {2\pi }
\sqrt{\vec k^2  + N^{8} \, {(k_1 N^3+k_2 N^2 + k_3 N + k_4 )^2
\over 1+N^2+N^4+N^6}}\,.
\ee
We now introduce new integers $m_1,m_2,m_3,m_4$ via
\be\label{echange}
k_4 = m_4- N m_3, \quad k_3 = m_3 - N m_2, \quad k_2 = m_2 - N m_1, \quad k_1 = m_1\, ,
\ee
or equivalently,
\be
m_1=k_1, \qquad m_2=k_2+Nk_1, \qquad m_3 = k_3 + Nk_2 + N^2 k_1, \qquad
m_4=k_4 + N k_3 + N^2 k_2 + N^3 k_1\, ,
\ee
so that 
\be\label{enewlength}
L(\vec k) = {2\pi }\, \sqrt{ N^2(m_1^2+m_2^2 + m_3^2 + m_4^2 + a_{ij} m_i m_j)
+k_5^2+k_6^2}, \qquad a_{ij}
=\OO(N^{-1})\, . 
\ee
The transformation \refb{echange} is unimodular, so any cycle on $T^6$ will correspond to
some choice of  integers $m_1,m_2,m_3,m_4,k_5,k_6$.
\refb{enewlength} now shows that $T^6$ can be considered as product of six circles, four each
of radius $N$ and two each of radius 1. 
This produces a near horizon geometry where
\be \label{e9.15a}
\wt R_H\sim N, \qquad R_H \sim 1, \qquad g^{(4)}_H\sim N^{-2}\, .
\ee
We see from \refb{eftheory3} that this would give $g_B\sim 1$, $R_B\sim 1$ and $\wt R_B\sim N$.
Therefore we have not yet established the existence of a region where $R_B$ is large. 
This can be rectified by
making a continuous SL(2,$\RRR)$ 
duality transformation of the solution.\footnote{A continuous duality 
transformation acting on an electrically charged black hole produces black holes
carrying both electric and magnetic charges. Generically these charges do not satisfy quantization law,
but for large charges we can simply pick the nearest allowed values of the charges after duality
rotation without significantly affecting the near horizon geometry.}
This leaves the $T^6$ moduli unchanged, and changes the axion dilaton
moduli $\tau$ whose imaginary part is $(g^{(4)}_H)^{-2}$. Since we also want to
preserve the asymptotic value of $\tau$, we are only allowed an $SO(2)$ subgroup
of the SL(2,$\RRR)$ group. If for definiteness we take the asymptotic value of $\tau$ to be
$i$ then the allowed transformation is 
$\tau\to (\tau\cos\theta+\sin\theta)/(-\tau\sin\theta+\cos\theta)$. So if we denote the near
horizon value of $\tau$ corresponding to \refb{e9.15a}
by $i\Lambda$ for some large $\Lambda$,
then after the duality transformation we get
\be
\tau = {i\Lambda\cos\theta+\sin\theta\over -i\Lambda\sin\theta+\cos\theta}
= i \, {\Lambda \over \Lambda^2\sin^2\theta+\cos^2\theta} + {(1-\Lambda)\cos\theta\sin\theta
\over  \Lambda^2\sin^2\theta+\cos^2\theta} \, .
\ee
If we take $\sin\theta\sim \Lambda^{-\eta}$  with ${1\over 2}<\eta<{1}$, then we get
\be
g^{(4)}_H= \tau_2^{-1/2} \sim \Lambda^{{1\over 2}-\eta }
\ee
Using the fact that $\Lambda\sim N^4$ in this case, we get the configuration
\be
\wt R_H\sim N, \qquad R_H \sim 1, \qquad g^{(4)}_H\sim N^{2-4\eta}\, .
\ee
\refb{eftheory3} now gives
\be
g_B\sim 1, \qquad \wt R_B\sim N^{2\eta-1}, \qquad R_B \sim N^{2-2\eta}\, .
\ee
Hence for ${1\over 2}<\eta < 1$, both 
$R_B$ and $\wt R_B$ are large, and $g_B$
remains finite.

This shows that  we can find
a black hole solution carrying appropriate 
electric and magnetic charges for which in the region
\refb{e5.12} space-time is best described as an F-theory background with large size of
the base $B=T^2/\ZZZ_2$ and hence the $(s,r)$ seven branes present in the F-theory 
compactification will get exposed in this region. Furthermore, as discussed in
\cite{2502.07883,2503.00601,2506.13876}, 
by scaling the parameter $m$ and the coordinate $\rho$ by a large number we
can make the region in the 3+1 dimensional space-time,
where the F-theory description is valid, to have arbitrarily large size.

\subsection{End of the world $E_8$ branes from M-theory compactification}  \label{smtheory}

We consider now $E_8\times E_8$ heterotic string theory on $T^6$
with unbroken $E_8\times E_8$. This has a dual description as
M-theory compactified on $S^1/\ZZZ_2\times T^6$. $S^1/\ZZZ_2$ can be regarded as the
segment of a real line and we have end of the world 9-branes at the two ends of the
line segment where the $E_8$ gauge theories live\cite{9510209}. 
Our goal will be to explore what kind of
field configurations in the heterotic string theory on $T^6$ will produce large space-time
regions where the M-theory description is valid and the end of the world $E_8$ branes get
exposed to an asymptotic observer..

Let us denote by $R_M$ the size of the interval and by $\wt R_M$ the size of the circles
of $T^6$, measured in eleven dimensional Planck scale. 
For the eleven dimensional M-theory description to be valid, we need $R_M$ and $\wt R_M$
to be large. Now, this theory can also
be regarded as
heterotic string theory on $T^6$, with the ten dimensional 
heterotic string coupling $g_H$, the
size $\wt R$ of $T^6$ measured in the heterotic string metric and the four dimensional
heterotic string metric $g^{(4)}_H$ given by\cite{9510209},
\be \label{e521}
g_H=R_M^{3/2}, \qquad \wt R_H = R_M^{1/2}\, \wt R_M, \qquad g^{(4)}_H = 
g_H \wt R_H^{-3}
= \wt R_M^{-3}\, .
\ee
Our goal is to have $R_M$ and $\wt R_M$ large.   
To achieve this, we modify the solution described in
section \ref{sftheory} by taking,
\be
\vec n = (1, N^{-1}, N^{-2}, N^{-3},N^{-4},N^{-5}) / \sqrt{1+N^{-2}+N^{-4}
+N^{-6}+N^{-8}+N^{-10}}\, ,
\ee
for some large integer $N$.
Then using \refb{e9.6} we see that 
a cycle in $T^4$ that joins the point $\vec 0$ to $2\pi\vec k$
in the $(x^4,x^5,x^6,x^7,x^8,x^9)$ directions for some integers 
$k_1,k_2,k_3,k_4.k_5,k_6$, has
length
\be
L(\vec k)=2\pi \sqrt{\vec k^2 +  { 2m\over \rho} \, \sinh^2\alpha\, ( \vec k.\vec n)^2}
\ee
So for the choice
\be \label{e5.12a}
{ 2m\over \rho} \, \sinh^2\alpha\, = N^{12} , 
\ee
we get 
\be
L(\vec k) = {2\pi} \sqrt{\vec k^2  + N^{12}\,  { (k_1 N^5+k_2 N^4 + k_3 N^3 + k_4 N^2 + k_5 N
+k_6 )^2\over 1+N^2+N^4 +N^6+N^8+N^{10}}}\,. 
\ee
If introduce new integers $m_1,m_2,m_3,m_4,m_5,m_6$ via
\be\label{echangea}
k_1 = m_1,
\quad k_i = m_i - N m_{i-1}, \quad \hbox{for $2\le i\le 6$}\, ,
\ee
then 
\be\label{enewlengtha}
L(\vec k) = {2\pi \, N}\, \sqrt{ m_1^2+m_2^2 + m_3^2 + m_4^2 +m_5^2+m_6^2 
+ a_{ij} m_i m_j}, \qquad a_{ij}
=\OO(N^{-1})\, . 
\ee
\refb{enewlengtha} now shows that for large $N$,
$T^6$ can be considered as product of six circles, each
of radius $N$. 
This gives $\wt R_H \sim N$. Also for this background, \refb{e9.6} gives $g^{(4)}_H \sim N^{-3}$. 
However by making a continuous SL(2,$\RRR)$ duality transformation of the type described at the
end of section \ref{sftheory}, 
we can make $g^{(4)}_H \sim N^{3-6\eta}$ for ${1\over 2}<\eta<1$. This gives
\be
\wt R_H \sim N, \qquad g^{(4)}_H \sim N^{3-6\eta}\, .
\ee
\refb{e521} now gives
\be
\wt R_M\sim N^{2\eta-1}, \qquad R_M\sim N^{4(1-\eta)} \, .
\ee
Hence for ${1\over 2}<\eta<1$ we shall achieve large values of $R_M$ and $\wt R_M$.

This shows that an observer in the heterotic string theory on $T^6\times \RRR^{3,1}$ can study
M-theory on an interval and the end of the world $E_8$ branes to any degree of accuracy.

\subsection{D8-branes in type I$'$ string theory}

D8-branes are codimension one branes in ten dimensional type IIA string theory, but instead of
having the usual Minkowski vacua of type IIA string theory on two sides, it separates AdS
vacua of type IIA string theory known as the Romans supergravity\cite{romans}. 
While we do not have direct world-sheet construction of a string theory whose low energy
limit is given by Romans supergravity, the latter appears as part of the background in type IIA
string theory compactified on $S^1/\ZZZ_2$, with orientifold 8-planes at the two ends and 16 
D8-branes placed at various points along $S^1/\ZZZ_2$\cite{9510169}.
This description becomes more and more accurate in the limit when the size $R_A$ 
of the interval measured in the type IIA metric grows
but the string coupling remains finite or small.\footnote{As in the case of F-theory, the
type IIA string coupling varies over the interval $S^1/\ZZZ_2$ unless each of the orientifold 8-plane
has 8 D8-branes on it\cite{9510169}. When each orientifold plane has 8 D8-branes
near it, then the string coupling remains constant away from the orientifold D8-brane system
and the $g_A$ appearing in the various formulae\ below can be taken to be the value of the
string coupling in this region. In the limit where the D8-branes are on top of the orientifold planes,
the $SO(32)$ group is broken to $SO(16)\times SO(16)$. This is further broken to $U(1)^{16}$
when the D8-branes are pulled away from the orientifold planes and from each other.\label{fo6}} 
We shall call the $S^1/\ZZZ_2$ direction as the 9th direction.
We shall consider four dimensional string theory
by compactifying this further on a five dimensional 
torus of size of order $\wt R_A$ which we also take to
be large. We shall call these additional compact directions as 4,5,6,7,8 directions.
Our goal will be to first map this to a heterotic string compactification on $T^6$ via a
series of duality transformation and then identify a black hole solution whose near horizon 
geometry produces large $R_A$ and $\wt R_A$ with small but finite $g_A$. 

We first make a
T-duality along the 9-th direction to convert it to a type I theory, with the type I string coupling 
$g_I$ and the size $R_I$ of the 9-th
direction and size $\wt R_I$ of the 4,5,6,7,8 directions, measured in the type I metric, given by
\be
R_I \sim R_A^{-1}, \qquad g_I \sim g_A \, R_A^{-1}, \qquad \wt R_I \sim \wt R_A\, .
\ee
In Planck units, the sizes $r$ of the 9th direction and $\wt r$ of the 4,5,6,7,8 directions are:
\be
\wt r \sim  \wt R_I g_I{}^{-1/4}\sim 
\wt R_A R_A^{1/4} g_A^{-1/4} , \qquad r \sim  R_I g_I{}^{-1/4} 
\sim  R_A^{-3/4} \, g_A^{-1/4} \, .
\ee
We now go to the heterotic description using the heterotic type I duality. The heterotic coupling 
$g_H$ is
given by $1/g_I$ up to a constant of proportionality and the radii of the circles measured in
Plank units remain unchanged. Denoting by $R_H$ and $\wt R_H$ the sizes of the 9-th direction
and the 4,5,6,7,8 directions measured in the heterotic string metric, we get
\be \label{eftheory1a}
g_H 
\sim  1/ g_I \sim  g_{A}^{-1}R{}_{A}, \qquad
\wt R_H \sim  \wt r \, g_H{}^{1/4} \sim  \wt R{}_{A}  R^{1/2}_{A}{} g_{A}^{-1/2}\, , \quad
R_H \sim  r \, g_H^{1/4} \sim  R_A^{-1/2} g_{A}^{-1/2}  \, .
\ee
From this we get the four dimensional heterotic string coupling to be
\be\label{eftheory2a}
g^{(4)}_H\sim  g_H \wt R_H^{-5/2} R_H^{-1/2} \sim  g_{A}^{1/2} \wt R_{A}^{-5/2}\, .
\ee
We see from \refb{eftheory1a} that in the limit of large $R_A$, $\wt R_A$ with fixed $g_A$,
$\wt R_H$ becomes large but $R_H$ becomes small. To bring this to a form analyzed earlier where
all circles have large size in the heterotic description, 
we make a further T-duality transformation 
along the 9-direction sending $R_H$ to $R_H^{-1}$ but
leaving $\wt R_H$ and $g^{(4)}_H$ unchanged. This maps the SO(32) heterotic string theory
to $E_8\times E_8$ heterotic string theory, both broken 
(approximately) to $SO(16)\times SO(16)$ by a
Wilson line along the 9-th direction. 
Denoting the resulting parameters by $~'$
we get
\be \label{eftheory4a}
\wt R_H' \sim   \wt R{}_{A}  R^{1/2}_{A}{} g_{A}^{-1/2}, \quad
R_H' \sim    R_A^{1/2} g_{A}^{1/2}, \quad
g^{(4)\prime}_H\sim    g_{A}^{1/2} \wt R_{A}^{-5/2}\, .
\ee
Finally, inverting \refb{eftheory4a}  we get
\be \label{eftheory3a}
g_A\sim \left(g^{(4)\prime}_H\right)^{-1/2} \wt R'_H{}^{-5/4} R_H'{}^{5/4}, 
\quad \wt R_A\sim \left(g^{(4)\prime}_H\right)^{-1/2} \wt R_H'{}^{-1/4} R_H'{}^{1/4} , 
\quad R_A \sim \left(g^{(4)\prime}_H\right)^{1/2} \wt R_H'{}^{5/4} R_H'{}^{3/4} \, .
\ee

Our goal will be to produce the configuration \refb{eftheory4a} with large $R_A$ and $\wt R_A$
but finite $g_A$ near the horizon of an
appropriate black hole.
\refb{eftheory4a} shows that in order to get large $R_A$ and $\wt R_A$ with fixed $g_A$
we need both $\wt R'_H$ and $R'_H$ to be large but $\wt R'_H$ should be large compared
to $R'_H$ and $g^{(4)\prime}_H$ should be small. To achieve this we consider the solution
considered in section \ref{smtheory} but replace \refb{e5.12a} by
\be \label{e5.12aa}
{ 2m\over \rho} \, \sinh^2\alpha\, = N^{12-\gamma}, \qquad 0<\gamma<2\, . 
\ee
In this case \refb{enewlengtha} is replaced by
\be\label{enewlengthaa}
L(\vec k) = {2\pi \, N}\, \sqrt{ m_1^2+m_2^2 + m_3^2 + m_4^2 +m_5^2+ N^{-\gamma}m_6^2 
+ a_{ij} m_i m_j}, \quad a_{ij} 
=\cases{\OO(N^{-1}) \, \hbox{for $i\ne j$}\cr 
 \OO(N^{-2}) \, \hbox{for $i= j$}} 
\, .
\ee
This gives
\be
\wt R_H'\sim N, \qquad R_H'\sim N^{1-\gamma/2}\, .
\ee
Also we have $g^{(4)}_H{}' \sim N^{-3+{1\over 4}\gamma}$, but 
with the help of a continuous duality rotation, we can convert it to
\be \label{enewlengthaab}
g^{(4)}_H{}' \sim N^{(-3+{1\over 4}\gamma)(2\eta-1)}\, ,
\ee
as before. Requiring that $g_A$ computed from
\refb{eftheory4a} is of order unity, now gives
\be
2\eta-1 = {5\gamma\over 12 - \gamma}\, .
\ee
Substituting this in \refb{enewlengthaa}, \refb{enewlengthaab} and using 
\refb{eftheory4a}, we now get
\be
R_A = N^{2-\gamma}, \qquad \wt R_A\sim N^{\gamma/2}\, .
\ee
Therefore, for $0<\gamma<2$, we have large $R_A$ and $\wt R_A$ as desired.

This shows that an observer in heterotic string theory on $T^6\times \RRR^{3,1}$ can not only
study D8-branes but also Romans supergravity in $\RRR^{9,1}$ to any desired degree of
accuracy. There is however a limit on the cosmological constant of the Romans supergravity
that can be studied this way since we only have a limited number of D8-branes.

\sectiono{Flat codimension one branes} \label{sflat}

We have seen that codimension one branes are difficult to study since macroscopic
loop of such branes are hidden behind the event horizon. In special cases when they are
part of compactification, one may be able to study them by producing a large region in
space-time where the compact directions tangential and transverse to the brane become
large, exposing the brane to an asymptotic observer. Examples of this for BPS branes
were discussed in section \ref{s5}. However this generally does not work for 
non-supersymmetric branes, {\it e.g.} there is no known compactification in which the
conjectured 8-brane separating type IIA and type IIB vacua appears in the vacuum. 

In this section we shall explore if flat codimension one branes can be studied by
an asymptotic observer. While flat codimension one branes have infinite energy and
hence cannot be regarded as a state of the theory, we could imagine a scenario where
they are present as part of the background, breaking translation invariance along
directions transverse to the brane. We shall call the two sides of the brane as 
$A$ and $B$ and also label the vacua on the two sides as $A$ and $B$. 
The question that we shall be addressing
is the following. Can an observer residing on side $A$ of the brane send an apparatus
to the side $B$ which performs an experiment over a macroscopic time scale and then
sends the information back to the original observer residing on side $A$? 
If this is the case then we can say
that the observer at infinity can measure properties of the vacua on both sides
of the brane to arbitrary accuracy, and the brane achieves the purpose of unifying the
vacua on two sides.

For this analysis we need to postulate some of the properties of the brane.
If the brane sources any other
scalar field $\phi$ then away from the brane we shall have a background 
in which $\phi$ depends on the coordinate $z$ transverse to the brane.
This will produce non-zero energy momentum tensor and the geometry
away from the brane will not be vacuum of the theories $A$ and $B$ as desired.
To avoid this situation we shall assume that the brane acts as a source of only the
gravitational field and no other massless field.\footnote{This would hold automatically if the
brane separates two Minkowski vacua in each of which all moduli have been stabilized.}
Put another way, we assume that sufficiently far away from the brane all fields other than
the metric become constant and
the tension of the brane is measured by integrating the energy momentum
tensor to sufficient distance in the transverse direction where variation of the other fields
have died down.
We shall further assume that the component
of the energy-momentum tensor along the brane is proportional to the induced metric
on the brane and the component of the energy momentum tensor transverse to the
brane vanishes.

The metric produced by such a brane was found in \cite{vilenkin}. For a $(D-1)$-brane in
a $(D+1)$ dimensional theory, it takes the form:
\be
ds^2 = e^{-2|\sigma|/\kappa} \left[-d\tau^2 + d\sigma^2 
+ e^{2 \tau/\kappa} \sum_{i=1}^{D-1} (dy^i)^2\right]\, .
\ee
Here $\tau$ is the time coordinate, $\sigma$ 
is the coordinate transverse to the brane, $y^i$ are
the coordinates tangential to the brane and $\kappa$ is a constant related to the brane tension. 
It was also 
shown in \cite{vilenkin} that away
from the location of the brane the space-time metric is flat. 
Hence we can go to a coordinate
system in which the metric away from the brane takes the form of the Minkowski metric.
The coordinate transformation is given by
\be
 t = \kappa e^{|\sigma| / \kappa} \sinh{\tau\over \kappa} +{1\over 2\kappa} e^{\tau/\kappa} \vec y^2,
\qquad z= \kappa e^{|\sigma| / \kappa} \cosh{\tau\over \kappa} - {1\over 2\kappa} 
e^{\tau/\kappa} \vec y^2,
\qquad
x^i = e^{(\tau+|\sigma|)/\kappa} y^i\, .
\ee
The metic in this new coordinate system takes the form:
\be
ds^2 = -dt^2 + dz^2 + \sum_{i=1}^{D-1} dx^i dx^i\, .
\ee
However in this coordinate system the brane moves along the transverse direction with uniform
acceleration, with the acceleration determined by the tension of the brane.  
We shall denote by $t$ the time and by $z$ the coordinate transverse
to the brane in this frame.
Then the trajectory
of the $\vec y=0$ point on the brane in the $z$-$t$ plane takes the form
\be \label{etraj}
t = \kappa \, \sinh{\tau\over \kappa}, \qquad z = \kappa\cosh{\tau\over \kappa}\, ,
\ee
with the acceleration of the brane given by $\kappa^{-1}$. 
Physically this is a reflection of the
fact that the brane exerts a constant repulsive gravitational force that makes an object outside
accelerate away from the brane. So in the inertial frame of the object the brane will accelerate
away from the object.

\def\figone{

\def\JPicScale{0.8}
\ifx\JPicScale\undefined\def\JPicScale{1}\fi
\unitlength \JPicScale mm
\begin{picture}(115,80)(0,0)
\linethickness{0.3mm}
\put(10,40){\line(1,0){100}}
\linethickness{0.3mm}
\put(60,0){\line(0,1){80}}
\linethickness{0.3mm}
\linethickness{0.3mm}
\multiput(20,0)(0.12,0.12){667}{\line(1,0){0.12}}
\linethickness{0.3mm}
\multiput(20,80)(0.12,-0.12){667}{\line(1,0){0.12}}
\linethickness{0.7mm}
\qbezier(100,80)(92.18,69.58)(87.97,63.56)
\qbezier(87.97,63.56)(83.76,57.55)(82.5,55)
\qbezier(82.5,55)(81.2,52.41)(80.59,50)
\qbezier(80.59,50)(79.99,47.59)(80,45)
\qbezier(80,45)(80,42.41)(80,40)%
\qbezier(80,40)(80,37.59)(80,35)%
\qbezier(80,35)(79.99,32.41)(80.59,30)
\qbezier(80.59,30)(81.2,27.59)(82.5,25)
\qbezier(82.5,25)(83.78,22.42)(86.19,18.81)
\qbezier(86.19,18.81)(88.59,15.2)(92.5,10)
\qbezier(92.5,10)(96.41,4.78)(98.22,2.38)
\qbezier(98.22,2.38)(100.02,-0.03)(100,0)

\linethickness{0.3mm}
\put(80,20){\line(0,1){40}}

\put(65,80){\makebox(0,0)[cc]{$t$}}

\put(115,40){\makebox(0,0)[cc]{$z$}}

\put(70,38){\makebox(0,0)[cc]{$\kappa$}}

\put(79,17){\makebox(0,0)[cc]{$P$}}

\put(79,63){\makebox(0,0)[cc]{$Q$}}

\end{picture}

}

\begin{figure}
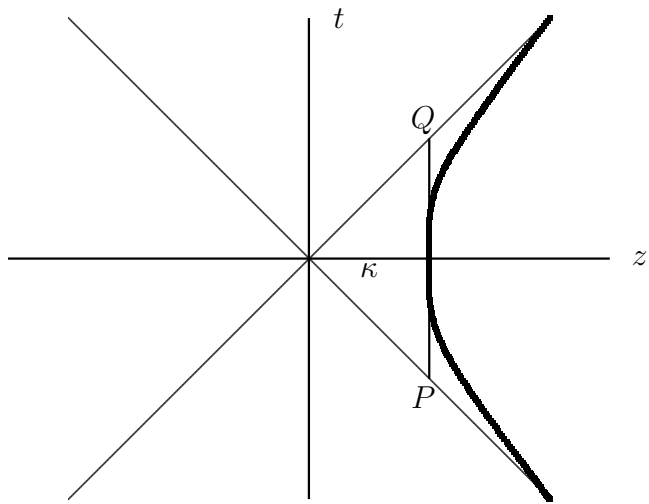


\begin{center}

\figone

\end{center}

\caption{This figure illustrates the maximum period during which an apparatus coming from side
$A$ of the brane to side $B$ and then returning to $A$
can be in the inertial frame. 
The thick curve is the trajectory of the brane as viewed from side $B$, which is on the
left side of the thick curve. The line segment $PQ$ denotes the
part of the trajectory of the apparatus on side $B$ along which the apparatus is in free fall. 
The view from side $A$ has not been
shown in this figure but it is of the same form as from side $B$. \label{figone}}

\end{figure}

Let us now examine the task at hand. Let the original observer be on side $A$ of the
brane. If it is within the past light cone of the brane, it can send an apparatus
across the brane to the other side $B$. After reaching side $B$, the apparatus can undergo
successive motion of acceleration and free fall any number of times 
and eventually send back the information gathered to the observer on the side $A$. The actual 
experiment has to be done during the free fall part of the trajectory since this is the
inertial frame in which side $B$ will appear to be in its vacuum. Thus the question we
want to address is: can we make a free fall phase last for arbitrarily long time, measured
by the clock in the apparatus?

To answer this question we can assume, without loss of generality, that the free fall phase
in which the proper time of the apparatus is maximum
corresponds to constant $z$ coordinate. This can be achieved by going to a boosted frame
in which the trajectory \refb{etraj} remains invariant but the inertial apparatus is at rest.
In that case the proper time measured by the inertial observer will be given by the lapse
of $t$ along the trajectory. The maximum possible time interval for which such an observer
can exist is seen to be $2\kappa$ according to Fig.~\ref{figone}. Since $\kappa$ is fixed by the
tension of the brane, we see that this time cannot be made arbitrarily large. Therefore we
conclude that the existence of such a brane does not allow an observer on side $A$ to make
measurement on side $B$ to arbitrary accuracy.

We end by noting that if the observer itself is allowed to cross the brane and go to the other side,
then it is possible for the world-line of the observer to have infinitely large inertial segments on both
sides of the brane.
This can be seen from Fig.~\ref{figone} by regarding this as representing the side $A$ instead
of side $B$. There are clearly infinitely long time-like geodesics (e.g. vertical lines)  
that intersects the 
trajectory of the brane. Such a trajectory spends infinitely long period in $A$ and then crosses the
brane. Once it emerges on the side $B$, the allowed trajectories can again be studied from
Fig.~\ref{figone}, this time regarding this as the side $B$. Without loss of generality we can take
the point of emergence as the point where the trajectory intersects the real axis and we see that
a vertical trajectory emerging from this point can exists for infinite time 
as long as it does not intend to return to $A$. 
Such an observer will be able to make arbitrarily accurate measurements
on both sides. 
This will be in accordance with the recent proposal of tying the measurements
to the world-line of a single observer instead of to the asymptotic 
region\cite{2206.10780,2308.03663}. This however postulates the existence of a meta observer
who can exist on both sides of the brane, even though the elementary particles themselves
differ on the two sides. This should be distinguished from the case where one needs to only send
an apparatus to the other side, which may be achieved {\it e.g.} by an appropriate gravitational wave
signal that constructs the apparatus after crossing to the other side, performs the desired 
experiment and sends back the result via gravitational wave signal.

\bigskip

\noindent{\bf Acknowledgement:} 
I would like to thank Tom Banks, Oren Bergman, 
Raghu Mahajan and Nati Seiberg for useful discussion. I would also like to acknowledge
ChatGPT for providing useful information.
This work was supported by the ICTS-Infosys Madhava 
Chair Professorship
and the Department of Atomic Energy, Government of India, under project no. RTI4019.


\begin{thebibliography}{99}

\bibitem{1011.5120}
T.~Banks and N.~Seiberg,
``Symmetries and Strings in Field Theory and Gravity,''
Phys. Rev. D \textbf{83} (2011), 084019
doi:10.1103/PhysRevD.83.084019
[arXiv:1011.5120 [hep-th]].



\bibitem{9302038}
A.~Sen,
``SL(2,Z) duality and magnetically charged strings,''
Int. J. Mod. Phys. A \textbf{8}, 5079-5094 (1993)
doi:10.1142/S0217751X93002009
[arXiv:hep-th/9302038 [hep-th]].

\bibitem{9408083}
A.~Sen,
``Strong - weak coupling duality in three-dimensional string theory,''
Nucl. Phys. B \textbf{434} (1995), 179-209
doi:10.1016/0550-3213(94)00461-M
[arXiv:hep-th/9408083 [hep-th]].

\bibitem{9707217}
S.~Elitzur, A.~Giveon, D.~Kutasov and E.~Rabinovici,
``Algebraic aspects of matrix theory on T**d,''
Nucl. Phys. B \textbf{509} (1998), 122-144
doi:10.1016/S0550-3213(97)00622-6
[arXiv:hep-th/9707217 [hep-th]].

\bibitem{9712047}
M.~Blau and M.~O'Loughlin,
``Aspects of U duality in matrix theory,''
Nucl. Phys. B \textbf{525} (1998), 182-214
doi:10.1016/S0550-3213(98)00242-9
[arXiv:hep-th/9712047 [hep-th]].

\bibitem{9712075}
C.~M.~Hull,
``U duality and BPS spectrum of superYang-Mills theory and M theory,''
JHEP \textbf{07} (1998), 018
doi:10.1088/1126-6708/1998/07/018
[arXiv:hep-th/9712075 [hep-th]].

\bibitem{9712084}
N.~A.~Obers, B.~Pioline and E.~Rabinovici,
``M theory and U duality on T**d with gauge backgrounds,''
Nucl. Phys. B \textbf{525} (1998), 163-181
doi:10.1016/S0550-3213(98)00264-8
[arXiv:hep-th/9712084 [hep-th]].

\bibitem{9809039}
N.~A.~Obers and B.~Pioline,
``U duality and M theory,''
Phys. Rept. \textbf{318} (1999), 113-225
doi:10.1016/S0370-1573(99)00004-6
[arXiv:hep-th/9809039 [hep-th]].

\bibitem{1004.2521}
J.~de Boer and M.~Shigemori,
``Exotic branes and non-geometric backgrounds,''
Phys. Rev. Lett. \textbf{104}, 251603 (2010)
doi:10.1103/PhysRevLett.104.251603
[arXiv:1004.2521 [hep-th]].

\bibitem{1209.6056}
J.~de Boer and M.~Shigemori,
``Exotic Branes in String Theory,''
Phys. Rept. \textbf{532}, 65-118 (2013)
doi:10.1016/j.physrep.2013.07.003
[arXiv:1209.6056 [hep-th]].

\bibitem{2502.07883}
A.~Sen,
``Are Moduli Vacuum Expectation Values or Parameters?,''
[arXiv:2502.07883 [hep-th]].

\bibitem{2503.00601}
A.~Sen,
``How to Create a Flat Ten or Eleven Dimensional Space-time in the Interior of an Asymptotically Flat Four Dimensional String Theory,''
[arXiv:2503.00601 [hep-th]].

\bibitem{2506.13876}
A.~Sen,
``Decorating Asymptotically Flat Space-Time with the Moduli Space of String Theory,''
[arXiv:2506.13876 [hep-th]].

\bibitem{2501.17697}
T.~Banks,
``Old Ideas for New Physicists III: String Theory Parameters are NOT Vacuum Expectation Values,''
[arXiv:2501.17697 [hep-th]].

\bibitem{1810.05337}
D.~Harlow and H.~Ooguri,
``Constraints on Symmetries from Holography,''
Phys. Rev. Lett. \textbf{122} (2019) no.19, 191601
doi:10.1103/PhysRevLett.122.191601
[arXiv:1810.05337 [hep-th]].

\bibitem{1810.05338}
D.~Harlow and H.~Ooguri,
``Symmetries in quantum field theory and quantum gravity,''
Commun. Math. Phys. \textbf{383} (2021) no.3, 1669-1804
doi:10.1007/s00220-021-04040-y
[arXiv:1810.05338 [hep-th]].


\bibitem{9402032}
A.~Sen,
``Dyon - monopole bound states, selfdual harmonic forms on the multi - monopole moduli space, and SL(2,Z) invariance in string theory,''
Phys. Lett. B \textbf{329} (1994), 217-221
doi:10.1016/0370-2693(94)90763-3
[arXiv:hep-th/9402032 [hep-th]].

\bibitem{9602022}
C.~Vafa,
``Evidence for F theory,''
Nucl. Phys. B \textbf{469} (1996), 403-418
doi:10.1016/0550-3213(96)00172-1
[arXiv:hep-th/9602022 [hep-th]].

\bibitem{9602114}
D.~R.~Morrison and C.~Vafa,
``Compactifications of F theory on Calabi-Yau threefolds. 1,''
Nucl. Phys. B \textbf{473} (1996), 74-92
doi:10.1016/0550-3213(96)00242-8
[arXiv:hep-th/9602114 [hep-th]].

\bibitem{9603161}
D.~R.~Morrison and C.~Vafa,
``Compactifications of F theory on Calabi-Yau threefolds. 2.,''
Nucl. Phys. B \textbf{476} (1996), 437-469
doi:10.1016/0550-3213(96)00369-0
[arXiv:hep-th/9603161 [hep-th]].

\bibitem{9510209}
P.~Horava and E.~Witten,
``Heterotic and Type I string dynamics from eleven dimensions,''
Nucl. Phys. B \textbf{460} (1996), 506-524
doi:10.1201/9781482268737-35
[arXiv:hep-th/9510209 [hep-th]].

\bibitem{9510169}
J.~Polchinski and E.~Witten,
``Evidence for heterotic - type I string duality,''
Nucl. Phys. B \textbf{460} (1996), 525-540
doi:10.1016/0550-3213(95)00614-1
[arXiv:hep-th/9510169 [hep-th]].

\bibitem{romans}
L.~J.~Romans,
``Massive N=2a Supergravity in Ten-Dimensions,''
Phys. Lett. B \textbf{169} (1986), 374
doi:10.1016/0370-2693(86)90375-8

\bibitem{9605150}
A.~Sen,
``F theory and orientifolds,''
Nucl. Phys. B \textbf{475} (1996), 562-578
doi:10.1016/0550-3213(96)00347-1
[arXiv:hep-th/9605150 [hep-th]].

\bibitem{Dabholkar}
A.~Dabholkar, G.~W.~Gibbons, J.~A.~Harvey and F.~Ruiz Ruiz,
``Superstrings and Solitons,''
Nucl. Phys. B \textbf{340} (1990), 33-55
doi:10.1016/0550-3213(90)90157-9

\bibitem{9707207}
E.~Cremmer, H.~Lu, C.~N.~Pope and K.~S.~Stelle,
``Spectrum generating symmetries for BPS solitons,''
Nucl. Phys. B \textbf{520}, 132-156 (1998)
doi:10.1016/S0550-3213(98)00057-1
[arXiv:hep-th/9707207 [hep-th]].

\bibitem{vilenkin2}
A.~Vilenkin and E.~P.~S.~Shellard,
``Cosmic Strings and Other Topological Defects,''
Cambridge University Press, 2000,
ISBN 978-0-521-65476-0

\bibitem{0510033}
J.~Polchinski,
``Open heterotic strings,''
JHEP \textbf{09} (2006), 082
doi:10.1088/1126-6708/2006/09/082
[arXiv:hep-th/0510033 [hep-th]].


\bibitem{9711200}
J.~M.~Maldacena,
``The Large $N$ limit of superconformal field theories and supergravity,''
Adv. Theor. Math. Phys. \textbf{2} (1998), 231-252
doi:10.4310/ATMP.1998.v2.n2.a1
[arXiv:hep-th/9711200 [hep-th]].

\bibitem{9812012}
E.~Witten,
``AdS/CFT correspondence and topological field theory.,''
JHEP \textbf{12} (1998), 012
doi:10.1088/1126-6708/1998/12/012
[arXiv:hep-th/9812012 [hep-th]].

\bibitem{2208.09396}
O.~Bergman and S.~Hirano,
``The holography of duality in $ \mathcal{N} $ = 4 Super-Yang-Mills theory,''
JHEP \textbf{11} (2022), 069
doi:10.1007/JHEP11(2022)069
[arXiv:2208.09396 [hep-th]].

\bibitem{2212.05077}
M.~Dierigl, J.~J.~Heckman, M.~Montero and E.~Torres,
``IIB string theory explored: Reflection 7-branes,''
Phys. Rev. D \textbf{107} (2023) no.8, 086015
doi:10.1103/PhysRevD.107.086015
[arXiv:2212.05077 [hep-th]].

\bibitem{2302.00007}
A.~Debray, M.~Dierigl, J.~J.~Heckman and M.~Montero,
``The Chronicles of IIBordia: Dualities, Bordisms, and the Swampland,''
[arXiv:2302.00007 [hep-th]].

\bibitem{2507.21210}
J.~J.~Heckman, J.~McNamara, J.~Parra-Martinez and E.~Torres,
``GSO Defects: IIA/IIB Walls and the Surprisingly Stable $\mathrm{R}7$-Brane,''
[arXiv:2507.21210 [hep-th]].

\bibitem{2509.03573}
V.~Chakrabhavi, A.~Debray, M.~Dierigl and J.~J.~Heckman,
``Exploring Pintopia: Reflection Branes, Bordisms, and U-Dualities,''
[arXiv:2509.03573 [hep-th]].

\bibitem{2303.17623}
J.~Kaidi, K.~Ohmori, Y.~Tachikawa and K.~Yonekura,
``Nonsupersymmetric Heterotic Branes,''
Phys. Rev. Lett. \textbf{131} (2023) no.12, 121601
doi:10.1103/PhysRevLett.131.121601
[arXiv:2303.17623 [hep-th]].

\bibitem{2411.04344}
J.~Kaidi, Y.~Tachikawa and K.~Yonekura,
``On non-supersymmetric heterotic branes,''
JHEP \textbf{03} (2025), 211
doi:10.1007/JHEP03(2025)211
[arXiv:2411.04344 [hep-th]].

\bibitem{2507.12467}
J.~Gomis,
``The AdS/$\mathsf{C}$-$\mathsf{P}$-${\mathsf T}$ Correspondence,''
[arXiv:2507.12467 [hep-th]].


\bibitem{1909.10355}
J.~McNamara and C.~Vafa,
``Cobordism Classes and the Swampland,''
[arXiv:1909.10355 [hep-th]].

\bibitem{9611007}
A.~Kumar and C.~Vafa,
``U manifolds,''
Phys. Lett. B \textbf{396} (1997), 85-90
doi:10.1016/S0370-2693(97)00108-1
[arXiv:hep-th/9611007 [hep-th]].

\bibitem{0406102}
C.~M.~Hull,
``A Geometry for non-geometric string backgrounds,''
JHEP \textbf{10} (2005), 065
doi:10.1088/1126-6708/2005/10/065
[arXiv:hep-th/0406102 [hep-th]].

\bibitem{vilenkin}
A.~Vilenkin,
``Gravitational Field of Vacuum Domain Walls,''
Phys. Lett. B \textbf{133} (1983), 177-179
doi:10.1016/0370-2693(83)90554-3

\bibitem{2206.10780}
V.~Chandrasekaran, R.~Longo, G.~Penington and E.~Witten,
``An algebra of observables for de Sitter space,''
JHEP \textbf{02} (2023), 082
doi:10.1007/JHEP02(2023)082
[arXiv:2206.10780 [hep-th]].

\bibitem{2308.03663}
E.~Witten,
``A background-independent algebra in quantum gravity,''
JHEP \textbf{03} (2024), 077
doi:10.1007/JHEP03(2024)077
[arXiv:2308.03663 [hep-th]].



\end{thebibliography}
\end{document}